\documentclass[acmsmall]{acmart}
\AtBeginDocument{%
  }

\usepackage{comment}
\usepackage{float}
\usepackage{tikz}
\usetikzlibrary{arrows.meta}
\usepackage{amsmath}
\usepackage{graphicx}
\usepackage{xcolor}
\usetikzlibrary{calc, shapes.geometric, positioning}
\usetikzlibrary{quantikz2}
\usepackage{quantikz}
\usepackage{subcaption}
\usepackage{booktabs}
\usepackage{array}
\usepackage{multirow}   
\usepackage{url}  
\usepackage{makecell}
\usepackage{bm}
\providecommand{\ket}[1]{|#1\rangle}
\newsavebox{\mycircuitbox}
\usepackage{algorithm}
\usepackage{algpseudocode}
\newtheorem{theorem}{Theorem}[section]
\newtheorem{lemma}[theorem]{Lemma}
\newtheorem{proposition}[theorem]{Proposition}
\title{Gate-level Implementation and Resource Analysis of Lackadaisical Quantum Walk Search}

\author{Amit Saha}
\email{abamitsaha@gmail.com}
\affiliation{%
  \institution{DI-ENS, École Normale Supérieure, Université PSL, CNRS, INRIA, Paris}
  \city{Paris}
  \country{France}
}

\author{Debanjan Kola}
\email{debanjan410353@gmail.com}
\affiliation{%
  \institution{Ramakrishna Mission Vivekananda Educational and Research Institute, Belur, Howrah}
  \city{Howrah}
  \country{India}
}

\author{Nishanka Das}
\email{nishanka.das.rkmvcc@gmail.com}
\affiliation{%
  \institution{Ramakrishna Mission Vivekananda Centenary College, Rahara, Kolkata}
  \city{Kolkata}
  \country{India}
}

\author{Amlan Chakrabarti}
\email{acakcs@caluniv.ac.in}
\affiliation{%
 \institution{A K Choudhury School of Information Technology, University of Calcutta, Kolkata}
 \city{Kolkata}
 \state{West Bengal}
 \country{India}
 }

\renewcommand{\shortauthors}{Saha et al.}

\begin{document}

\begin{abstract}
Lackadaisical quantum walks (LQW) extend discrete-time quantum walks (DTQW) by introducing weighted self-loops, enabling improved spatial-search performance through controlled localization of the walker. Despite substantial theoretical progress, practical gate-level implementations suitable for quantum hardware remain largely unexplored, limiting evaluation under realistic architectural constraints, noise, and resource requirements. In this work, we present a gate-level implementation framework for lackadaisical quantum walk search. The proposed construction encodes the position and coin spaces into qubit registers, and realizes the walk dynamics through oracle, coin, and flip-flop shift operations. We validate the circuit by reproducing the expected search behavior for single and multiple marked vertices and by analyzing the effect of the self-loop weight on the success probability. We further evaluate the implementation under realistic noisy settings using superconducting hardware's noise models. Logical resource analysis shows that, for grids ranging from $8\times8$ to $64\times64$, the transpiled gate count increases from $3.63\times10^{5}$ to $4.38\times10^{6}$ and the circuit depth from $2.13\times10^{5}$ to $2.56\times10^{6}$. Finally, fault-tolerant resource estimates based on a surface-code model using the Microsoft Quantum Resource Estimator demonstrate the substantial space-time trade-off associated with magic-state production. The results establish a practical circuit-level pathway for implementing the LQW search and provide a basis for evaluating its performance. 
\end{abstract}

\begin{comment}

\begin{CCSXML}
<ccs2012>
 <concept>
  <concept_id>00000000.0000000.0000000</concept_id>
  <concept_desc>Do Not Use This Code, Generate the Correct Terms for Your Paper</concept_desc>
  <concept_significance>500</concept_significance>
 </concept>
 <concept>
  <concept_id>00000000.00000000.00000000</concept_id>
  <concept_desc>Do Not Use This Code, Generate the Correct Terms for Your Paper</concept_desc>
  <concept_significance>300</concept_significance>
 </concept>
 <concept>
  <concept_id>00000000.00000000.00000000</concept_id>
  <concept_desc>Do Not Use This Code, Generate the Correct Terms for Your Paper</concept_desc>
  <concept_significance>100</concept_significance>
 </concept>
 <concept>
  <concept_id>00000000.00000000.00000000</concept_id>
  <concept_desc>Do Not Use This Code, Generate the Correct Terms for Your Paper</concept_desc>
  <concept_significance>100</concept_significance>
 </concept>
</ccs2012>
\end{CCSXML}

\ccsdesc[500]{Do Not Use This Code~Generate the Correct Terms for Your Paper}
\ccsdesc[300]{Do Not Use This Code~Generate the Correct Terms for Your Paper}
\ccsdesc{Do Not Use This Code~Generate the Correct Terms for Your Paper}
\ccsdesc[100]{Do Not Use This Code~Generate the Correct Terms for Your Paper}
\end{comment}


\maketitle

\section{Introduction}
The rapid progress of quantum computing hardware has renewed interest in the practical realization of quantum algorithms \cite{nisqpreskil}. Foundational results such as Shor's factoring algorithm~\cite{shor1999polynomial} and Grover's search algorithm~\cite{grover1996fast} demonstrated that quantum computation can offer asymptotic advantages over the best-known classical approaches for specific computational tasks. In particular, Grover's algorithm reduces the query complexity of unstructured search from $O(N)$ to $O(\sqrt{N})$, motivating a broad class of quantum search procedures. Quantum walk search algorithms generalize this idea to structured search spaces such as graphs and lattices, where the evolution of a quantum walker can be exploited to amplify the probability of finding a marked vertex \cite{aaronson2003quantum}. Among these models, lackadaisical quantum walks introduce weighted self-loops at each vertex, allowing controlled localization of the walker and, with an appropriate choice of self-loop weight, improved spatial-search performance compared with standard quantum walk search algorithms~\cite{wong2018faster}.

\subsection{Motivation}

Recent studies have explored lackadaisical quantum walks in application areas such as quantum image processing, including edge detection and segmentation~\cite{giri2025quantum}, as well as cryptographic protocols based on modified walk dynamics~\cite{gibson2026nisq}. Despite substantial theoretical and application progress, a gap remains between the algorithmic promise of lackadaisical quantum walks and their practical realization on quantum hardware. Existing studies of lackadaisical quantum walk search have primarily focused on abstract evolution operators, asymptotic performance, and numerical analysis of the underlying quantum dynamics~\cite{giri2024edge, gibson2026nisq}. While these analyses establish the potential advantages of lackadaisical quantum walks, they generally assume that the corresponding walk operator can be implemented as an ideal unitary transformation. As a result, the gate-level synthesis of lackadaisical quantum walk search, including its position encoding, coin construction, oracle realization, and controlled shift dynamics, has not been fully addressed. This implementation gap is significant because lackadaisical quantum walks introduce additional circuit-level challenges beyond those encountered in standard discrete-time quantum walks. The inclusion of weighted self-loops enlarges the coin space and requires the walker to support both directional movement and a stay-at-place transition. Consequently, the coin operator must be embedded into a qubit-compatible Hilbert space, and the shift operator must be designed to correctly implement movement and self-loop behavior within a reversible quantum circuit. These requirements become increasingly demanding as the grid size grows, affecting the number of qubits, gate count, circuit depth, and the feasibility of execution on near-term quantum devices. A hardware-executable circuit construction is therefore necessary to evaluate the lackadaisical quantum walk search beyond ideal theoretical models. Such a construction enables numerical verification of the expected search dynamics, simulation under realistic noise models, and estimation of logical and physical resources required for scalable implementation. By developing and analyzing a complete gate-level framework, this work aims to bridge the gap between the theoretical formulation of lackadaisical quantum walks and their practical deployment on quantum computing platforms.

\subsection{Related Works}

Quantum search was first established as a fundamental application of quantum computation through Grover's algorithm~\cite{grover1996fast}, which finds a marked item in an unstructured search space with quadratic speedup compared to its classical counterpart. This quadratic improvement motivated the study of quantum search beyond unstructured databases, particularly on spatially structured search spaces such as grids, lattices, and other graphs. In these settings, quantum walks provide a natural framework for describing search dynamics, where the evolution of a quantum walker is used to amplify the probability of observing a marked vertex \cite{PhysRevA.67.052307}. For searching on two-dimensional periodic structures such as a torus, grid, or periodic square lattice with $N$ vertices, Aaronson and Ambainis~\cite{aaronson2003quantum} proposed a recursive algorithm that solves the problem in $\mathcal{O}(\sqrt{N}\log^2 N)$ time while achieving Grover-like speedup on higher-dimensional grids. Childs and Goldstone~\cite{childs2004spatial} later studied an alternative approach based on continuous-time quantum walks. Although the method does not provide an optimal speedup in two and three dimensions, it achieves a runtime of $\mathcal{O}(\sqrt{N}\log^{3/2}N)$ in four dimensions and $\mathcal{O}(\sqrt{N})$ in five or more dimensions. Subsequently, Ambainis \emph{et al.}~\cite{ambainis2004coins} introduced a discrete-time coined quantum walk algorithm that improved the two-dimensional search complexity to $(O(\sqrt{N}\log N))$, while retaining $(O(\sqrt{N}))$ scaling in higher dimensions. Building upon this work, Wong~\cite{wong2018faster} modified the discrete-time quantum walks of Ambainis \emph{et al.}~\cite{ambainis2004coins} by adding weighted self-loops at each vertex, thereby making the walker ``lazy''. This approach achieved a success probability close to unity in $\mathcal{O}(\sqrt{N\log N})$ time with an appropriate choice of the self-loop weight, representing a $\sqrt{\log N}$ speedup over the algorithm of Ambainis \emph{et al.}~\cite{ambainis2004coins}. Further studies have analyzed the role of the self-loop weight, the behavior under different numbers and arrangements of marked vertices, and the theoretical properties of the corresponding walk dynamics~\cite{wong2016faster, hoyer2017efficient}. On the other hand, the sensitivity of the walk's evolution to its initial parameters and self-loop weights has been explored in quantum cryptography, inspiring novel protocols for secure communication and quantum hash functions \cite{gibson2026nisq, li2013discrete}. Furthermore, the ability of the model to quickly cover the space of possible states has attracted considerable interest from optimization and finance, with LQW-derived methods being developed for such computationally difficult tasks as portfolio optimization and solving combinatorial problems \cite{qiang2024review, marsh2019quantum}. 

In parallel, several works have investigated circuit constructions and experimental demonstrations of standard quantum walks on near-term quantum platforms. Gate-level implementations of standard discrete-time quantum walks and quantum walk search have been studied with attention to qubit requirements, circuit depth, and suitability for noisy intermediate-scale quantum devices~\cite{zaidi2021experimental, razzoli2024efficient}. Frameworks such as IBM Quantum and Qiskit have further enabled the simulation and execution of such circuits on superconducting quantum processors~\cite{ibmquantum, Qiskit}. However, these implementations primarily address standard discrete-time quantum walks and do not provide a complete gate-level realization of lackadaisical quantum walk search, where the weighted self-loop requires an enlarged coin space, qubit-compatible embedding, and controlled shift operations that preserve the intended walk dynamics.

Therefore, while the theoretical advantages of lackadaisical quantum walks have been well studied, their practical implementation as executable quantum circuits remains underdeveloped. This work addresses that gap by presenting a complete circuit-level construction of lackadaisical quantum walk search, together with validation under ideal and noisy settings and an analysis of the logical and physical resources required for scalable implementation.

\subsection{Contributions}

The main contributions of this work are summarized as follows:
\begin{itemize}

\item We present explicit qubit-register gate-level construction of lackadaisical spatial search on a generic $2^n \times 2^n$ periodic grid.
%\item We present, to the best of our knowledge, the first complete gate-level implementation framework for lackadaisical quantum walk search on a generic two-dimensional grid, suitable for simulation and execution within superconducting quantum computing platforms.

\item We construct the required circuit components of the lackadaisical quantum walks, including position register initialization, embedding of the five-dimensional lackadaisical coin into a three-qubit unitary representation, phase-oracle implementation, and controlled flip-flop shift operations.

\item Through simulation, we validate the proposed circuit by reproducing the expected search behavior for both single and multiple marked vertices under ideal noiseless conditions.

\item We analyze the effect of the self-loop weight $\ell$ on the success probability, demonstrating how the lackadaisical parameter influences amplitude localization and search performance.

\item We evaluate the proposed implementation under noisy quantum-circuit simulations using superconducting device-inspired noise models, thereby assessing the robustness of the lackadaisical quantum walk search circuit in realistic settings.

\item We investigate strategies for improving the search performance in noisy settings by applying noise-mitigation techniques and by varying the self-loop weight $\ell$, and we compare the resulting success probabilities with the corresponding unmitigated noisy simulations.

\item We provide a resource analysis of the proposed construction, including logical circuit requirements and physical-resource estimates under a fault-tolerant implementation model based on surface code \cite{Fowler_2012}.

\end{itemize}

\begin{comment}

\subsection{Contribution}
\begin{itemize}
    \item This is the first time of its kind that the lackadaisical quantum walks have been implemented on a superconducting device.
    \item We validate all the theoretical properties of lackadaisical quantum walks through physical implementation. (one-dimensional line)
    \item We show the simulation of searching the single and multiple marked states through lackadaisical quantum walks.
    \item We further implement the lackadaisical search quantum walks on a 2D grid under the generic noise model.
    \item We apply a noise mitigation technique to get a better success probability for lackadaisical quantum search walks.
\item We analyze the resources for the circuit construction of lackadaisical quantum walks with surface code.
    
\end{itemize}

\end{comment}
\subsection{Organization}
This article is organized as follows. 
Section~\ref{sec:prelim} introduces the necessary preliminaries, including an overview of quantum circuits along with the foundational concepts of the lackadaisical quantum walks algorithm. 
Section~\ref{sec:impl} presents the circuit implementation, covering the 
circuit architecture, initialization, coin operator construction, and the 
flip-flop shift operator. 
Section~\ref{NR} provides numerical verification of the circuit, 
addressing three cases: single and multiple marked states in an ideal 
noiseless medium, behavior under noise, and the effect of noise mitigation techniques along with variation of the self-loop weight $\ell$. 
Section~\ref{sec:resource} details the resource analysis, comprising 
logical and physical cost estimates. 
Finally, Section~\ref{sec:conclusion} concludes with a summary and 
discussion of future directions.
\section{Preliminaries}\label{sec:prelim}
This section introduces the basic concepts required for the circuit-level implementation of lackadaisical quantum walks search. We first review discrete-time quantum walks, then describe their lackadaisical extension through weighted self-loops, and finally summarize the quantum-circuit components used in the proposed implementation.

\subsection{Discrete-time quantum walks}
\label{sec:dtqw}
A discrete-time quantum walk \cite{kempe2003quantum, 10.1145/380752.380758} is the quantum analogue of a classical random walk in which the evolution is governed by unitary operations. The state space consists of a position space and a coin space. A one-dimensional discrete-time quantum walk on the integer line $\mathbb{Z}$
operates on the composite Hilbert space
$\mathcal{H} = \mathcal{H}_P \otimes \mathcal{H}_C$,
where $\mathcal{H}_C = \mathrm{span}\{\ket{\leftarrow},\ket{\rightarrow}\}$ is the
two-dimensional coin space, where $\ket{\leftarrow}$ and $\ket{\rightarrow}$ denote movement to the left and right, respectively and
$\mathcal{H}_P = \mathrm{span}\{|x\rangle : x \in \mathbb{Z}\}$ is the
position space. Each step consists of a coin toss followed by a
conditional position shift. A common choice of coin in one dimension is the Hadamard operator:
\begin{equation}
    H = \frac{1}{\sqrt{2}}
    \begin{pmatrix} 1 & 1 \\ 1 & -1 \end{pmatrix}
\end{equation}
and the shift operator moves the walker left or right according to the
coin state:
\begin{equation}
    S = \ket{\leftarrow}\langle L| \otimes \sum_{x}|x{-}1\rangle\langle x|
      + \ket{\rightarrow}\langle R| \otimes \sum_{x}|x{+}1\rangle\langle x|.
\end{equation}
One complete step is the unitary $U = S(H \otimes I_P)$, where $I_P$ is the identity operator on the position space, so after $t$
steps the state is $|\psi_t\rangle = U^t|\psi_0\rangle$.
\subsection{Lackadaisical quantum walks}
\label{sec:lqw_def}

The lackadaisical quantum walks, introduced
by~\cite{wong2015grover}, extend the standard DTQW by adding a
weighted self-loop of weight $\ell \geq 0$ to every vertex.
This allows the walker to remain at its current position during a walk step.

\subsubsection{Lackadaisical quantum walks on a line}
\label{sec:lqw1d}

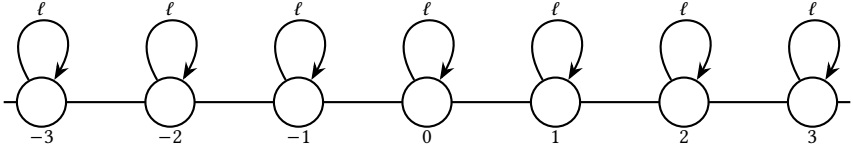
\begin{figure}[!h]
\centering
\begin{tikzpicture}[
    vertex/.style={circle, draw, thick,
                   minimum size=0.65cm, fill=white},
    >=Stealth
]
\def\s{1.7}

\draw[thick] ({-3*\s - 0.5}, 0) -- ({3*\s + 0.5}, 0);

\foreach \x in {-3,-2,-1,0,1,2,3}{
    \node[vertex] (v\x) at ({\x*\s}, 0) {};
    \node[below=7pt, font=\footnotesize] at ({\x*\s}, 0) {$\x$};
}

\foreach \x in {-3,-2,-1,0,1,2,3}{
    \draw[->, thick] (v\x)
        to[out=120, in=60, looseness=9]
        node[pos=0.5, above, font=\footnotesize,
             inner sep=2pt] {$\ell$}
        (v\x);
}
\end{tikzpicture}
\caption{A one-dimensional lackadaisical quantum walk on a line. }
\label{fig:lqw1d}
\end{figure}

For a one-dimensional lackadaisical quantum walk which is illustrated in Figure~\ref{fig:lqw1d} on a line, the coin space is enlarged from two directions to three:
\begin{equation}
\mathcal{H}_C^{(1D)}
=
\mathrm{span}
\left\{
\ket{\leftarrow}, \ket{\rightarrow}, \ket{\circlearrowleft}
\right\},
\end{equation}
where $\ket{\circlearrowleft}$ denotes the self-loop state. The corresponding weighted coin state is
\begin{equation}
|s_C^{(1D)}\rangle
=
\frac{1}{\sqrt{2+\ell}}
\left(
\ket{\leftarrow} + \ket{\rightarrow} + \sqrt{\ell}\ket{\circlearrowleft}
\right).
\end{equation}
The associated Grover-type coin operator is
\begin{equation}
C^{(1D)}
=
2|s_C^{(1D)}\rangle\langle s_C^{(1D)}|
-
I_3 .
\end{equation}
When $\ell=0$, the self-loop has no contribution and the walk reduces to the standard discrete-time quantum walk. As $\ell$ increases, the probability amplitude associated with remaining at the current vertex becomes more significant.

\begin{comment}

For a one-dimensional lackadaisical quantum walk, the coin space is enlarged from two directions to three: $\mathcal{H}_C^{(1D)} = \mathrm{span}\{\ket{\leftarrow},\ket{\rightarrow},|\circlearrowleft\rangle\}$, where $|\circlearrowleft\rangle$ denotes the self-loop direction.
The coin operator is the generalized Grover diffusion 
matrix~\cite{grover1996fast, portugal2013quantum},

\begin{equation}
    G = \frac{2}{2 + \ell}\,\mathbf{1}\mathbf{1}^{\!\top} - I.
    \label{eq:grover_coin_1d}
\end{equation}

Here, $\mathbf{1}$ denotes the all-ones column vector, and $I$ is the identity operator of the corresponding coin dimension. The shift operator is modified to include the self-loop state: the states $\ket{\leftarrow}$ and $\ket{\rightarrow}$ move the walker to positions $x-1$ and $x+1$, respectively, while the self-loop state leaves the walker at the same vertex. When $\ell=0$, the model reduces to the standard discrete-time quantum walk; as $\ell$ increases, the self-loop contribution becomes stronger and the walker becomes more localized. Figure~\ref{fig:lqw1d} illustrates a one-dimensional lackadaisical quantum walk on the integer line.

\end{comment}

\subsubsection{Lackadaisical quantum walks on two-dimensional grids}
\label{sec:lqw2d}

\begin{figure}[!h]
\centering
\begin{tikzpicture}[
    vertex/.style={circle, draw, thick,
                   minimum size=0.7cm, fill=white},
    marked/.style={circle, draw, thick,
                   double, double distance=2.5pt,
                   minimum size=0.7cm, fill=white},
    >=Stealth
]
\def\s{1.8}
\def\pad{0.5}

\foreach \r in {0,...,4}{
    \draw[thick] ({-\pad},{\r*\s}) -- ({4*\s+\pad},{\r*\s});
}
\foreach \c in {0,...,4}{
    \draw[thick] ({\c*\s},{-\pad}) -- ({\c*\s},{4*\s+\pad});
}

\foreach \r in {0,...,4}{
    \foreach \c in {0,...,4}{
        \pgfmathparse{(\r==2 && \c==2) ? 1 : 0}
        \pgfmathtruncatemacro{\ismarked}{\pgfmathresult}
        \ifnum\ismarked=1
            \node[marked] (n\r\c) at ({\c*\s},{\r*\s}) {};
        \else
            \node[vertex]  (n\r\c) at ({\c*\s},{\r*\s}) {};
        \fi
    }
}

\foreach \r in {0,...,4}{
    \foreach \c in {0,...,4}{
        \draw[->, thick] (n\r\c)
            to[out=160, in=110, looseness=9]
            node[pos=0.45, above left,
                 inner sep=1pt,
                 font=\footnotesize] {$\ell$}
            (n\r\c);
    }
}
\end{tikzpicture}
\caption{Lackadaisical quantum walks on a $5\times 5$ grid. Every vertex carries a self-loop of weight $\ell$. The double-circled vertex denotes the marked vertex.}
\label{fig:lqw2d}
\end{figure}
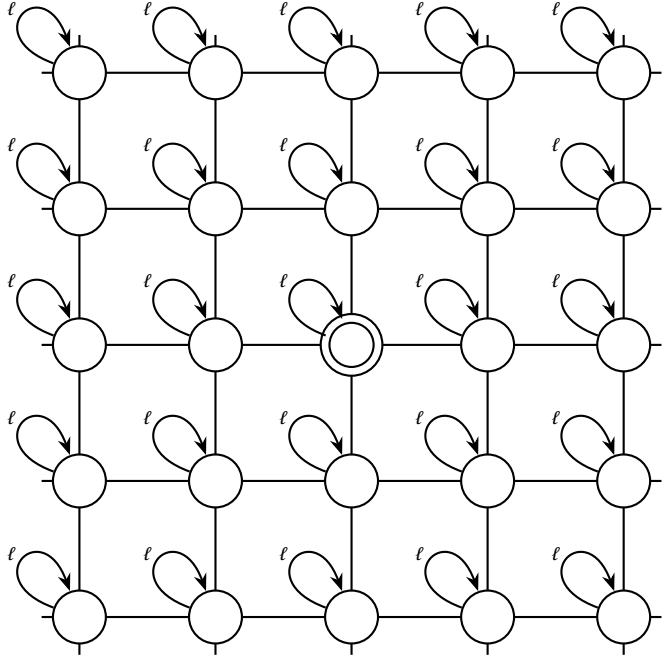

This work focuses on lackadaisical quantum walk search on a two-dimensional $L\times L$ grid with $N=L^2$ vertices, which is illustrated as an instance in Figure~\ref{fig:lqw2d}. In the standard two-dimensional discrete-time quantum walk, the walker can move in four directions: up, down, left, and right. The lackadaisical extension adds one self-loop direction, giving the five-dimensional coin space
\begin{equation}
\mathcal{H}_C^{(2D)}
=
\mathrm{span}
\left\{
\ket{\uparrow}, \ket{\downarrow}, \ket{\leftarrow}, \ket{\rightarrow}, \ket{\circlearrowleft}
\right\}.
\label{eq:lqwcoinspace}
\end{equation}
The weighted lackadaisical coin state is
\begin{equation}
|s_C^{(\ell)}\rangle
=
\frac{1}{\sqrt{4+\ell}}
\left(
\ket{\uparrow}
+
\ket{\downarrow}
+
\ket{\leftarrow}
+
\ket{\rightarrow}
+
\sqrt{\ell}\ket{\circlearrowleft}
\right).
\label{eq:lqwcoin}
\end{equation}
The corresponding five-dimensional Grover-type coin is
\begin{equation}
C_5^{(\ell)}
=
2|s_C^{(\ell)}\rangle\langle s_C^{(\ell)}|
-
I_5 .
\label{eq:lqwfivedimcoin}
\end{equation}

The shift operation used for two-dimensional lackadaisical quantum walks is the flip-flop shift. It moves the walker in the direction indicated by the coin state and then reverses the direction label. The self-loop state leaves the walker at the same vertex. This shift structure is important for preserving reversibility and is used in the circuit construction in Section~\ref{sec:impl}.

\begin{comment}

On a two-dimensional $\sqrt{N}\times\sqrt{N}$ grid with $N$ vertices, each internal vertex has degree $d = 4$. Adding a self-loop of weight
$\ell$ extends the coin space to five dimensions:
$\mathcal{H}_C =
\mathrm{span}\{|\uparrow\rangle,|\downarrow\rangle,
               |\leftarrow\rangle,|\rightarrow\rangle,
               |\circlearrowleft\rangle\}$.
The coin operator is the Grover diffusion matrix:

\begin{equation}
    G = \frac{2}{4 + \ell}\,J - I
    \label{eq:grover_coin_2d}
\end{equation}

where $J$ is the all-ones matrix of appropriate dimension.
One step of the 2D lackadaisical walks ~\cite{wong2018faster} is:

\begin{equation}
    U = S \cdot (G \otimes I_N)
    \label{eq:lqw_step}
\end{equation}

where $S$ is the flip-flop shift operator~\cite{portugal2016staggered} that moves the walkser along
the grid and reflects the coin direction upon arrival. The resulting
walks on the $\sqrt{N}\times\sqrt{N}$ grid is illustrated in
Figure~\ref{fig:lqw2d}.
\end{comment}

\subsection{Quantum search using lackadaisical quantum walks}
\label{sec:lqw_search}

In this paper, we discuss the search problem that is finding one or more marked vertices in the grid. Let $M$ denote the set of marked vertices. For a single marked vertex $w$, the phase oracle is
\begin{equation}
Q_w = I_N - 2|w\rangle\langle w| .
\label{eq:phase_oracle}
\end{equation}
For multiple marked vertices, the oracle is
\begin{equation}
Q_M =
I_N
-
2\sum_{w\in M}
|w\rangle\langle w| .
\label{eq:multiplemarked}
\end{equation}
This oracle flips the phase of the marked position states and leaves all other position states unchanged.

The lackadaisical quantum walk search begins from the initial state
\begin{equation}
|\Psi_0\rangle
=
|s_P\rangle \otimes |s_C^{(\ell)}\rangle,
\end{equation}
where
\begin{equation}
|s_P\rangle
=
\frac{1}{\sqrt{N}}
\sum_{v=0}^{N-1}
|v\rangle
\label{eq:initvertex}
\end{equation}
is the uniform superposition over all vertices. One step of the search walk is
\begin{equation}
U_{\mathrm{search}}^{(\ell)}
=
S_{\mathrm{ff}}
\left(I_N \otimes C_5^{(\ell)}\right)
\left(Q_M \otimes I_5\right),
\label{eq:lqwsearch}
\end{equation}
where $S_{\mathrm{ff}}$ is the flip-flop shift operator. Repeated applications of $U_{\mathrm{search}}^{(\ell)}$ amplify the probability of measuring a marked vertex.

The value of the self-loop weight $\ell$ controls the balance between movement across the grid and localization near marked vertices. If $\ell$ is too small, the self-loop has little effect; if it is too large, the walk becomes overly localized and movement is suppressed. For spatial search of one marked vertex or state on a two-dimensional grid, a commonly used optimal choice is $\ell = {4}/{N}$. In this work, we also study the effect of varying $\ell$, particularly under noisy circuit simulations, to examine whether parameter tuning can improve the observed success probability.

\begin{comment}

To perform a spatial search, a phase oracle marks the target vertex
$m \in V$:

\begin{equation}
    \mathcal{O} = I - 2|m\rangle\langle m|
    \label{eq:oracle}
\end{equation}

The modified walks operator $U' = U\cdot\mathcal{O}$ is applied
repeatedly, starting from the uniform superposition
$|\psi_0\rangle = \frac{1}{\sqrt{N}}\sum_{v=0}^{N-1}|v\rangle$.
The amplitude at the marked vertex grows with each step until a measurement reveals $m$ with high probability.

The self-loop weight $\ell$ governs the balance between exploration
and localization. Too small a value yields insufficient amplitude
concentration; too large a value suppresses movement entirely.
For spatial search on a $\sqrt{N}\times\sqrt{N}$ grid, the optimal self-loop weight is~\cite{wong2015grover, hoyer2017efficient}:

\begin{equation}
    \ell = \frac{4}{N}
    \label{eq:optimal_l}
\end{equation}

With this choice the lackadaisical quantum walks finds the marked vertex
with $\Omega(1)$ success probability in
$\mathcal{O}\left(\sqrt{N\log N}\right)$ steps~\cite{wong2016faster}, 
matching the known quantum lower bound~\cite{childs2004spatial}for search on two-dimensional grids. The initialization in the
uniform superposition and the Grover-diffusion coin together ensure
that the $\ell = 4/N$ self-loop provides precisely the localization
needed to concentrate the amplitude at the target without impeding global
exploration.

\end{comment}
\subsection{Quantum circuit model}

A quantum circuit represents a quantum algorithm as a sequence of gates applied to qubit registers. In this work, we follow the quantum circuit model that is used to realize the main components of a discrete-time quantum walk \cite{PhysRevA.79.052335}: state preparation, coin operation, conditional shift operation, phase oracle, and final measurement. In a similar way, in the proposed implementation of LQW on the grid, the position of the walker is encoded using two position registers, while the coin space is encoded using a separate three-qubit register. The main circuit components required for the lackadaisical quantum walk search are state preparation, the phase oracle, the embedded lackadaisical coin, the controlled flip-flop shift, and measurement. The elementary gates \cite{barenco1995elementary} used in the construction include single-qubit gates such as $H$, $X$, $Z$, $S$, $S^\dagger$, $T$, $T^\dagger$, together with controlled operations such as $CX$, and multi-controlled unitary gates. Since multi-controlled unitary gates cannot be implemented as a native hardware operation, Qiskit automatically decomposes the operator into native one- and two-qubit gates. This synthesis relies on established decompositions of multi-controlled operations into Clifford$+T$ elementary quantum gates~\cite{barenco1995elementary, PhysRevA.87.042302}. The matrix representations of the gates used in the implementation are summarized in Table~\ref{tab:quantum_gates_matrices} in Appendix \ref{app:quantum_gates_matrices}.

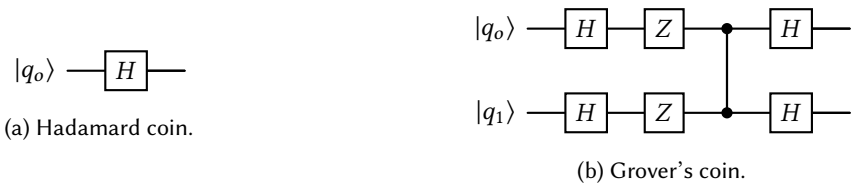
\begin{figure}[!h]
    \centering
    \begin{subfigure}[c]{0.35\textwidth}
        \centering
            \begin{quantikz}[column sep=0.5cm]
             \lstick{$\ket{q_o}$} & \gate{H} & \qw
        \end{quantikz}
        \caption{Hadamard coin.}
        \label{subfig:hadamard}
    \end{subfigure}
    \hfill
    \begin{subfigure}[c]{0.58\textwidth}
        \centering
        \begin{quantikz}[column sep=0.5cm, row sep=0.6cm]
            \lstick{$\ket{q_o}$} & \gate{H} & \gate{Z} & \ctrl{1}  & \gate{H} & \qw \\
            \lstick{$\ket{q_1}$} & \gate{H} & \gate{Z} & \ctrl{-1} & \gate{H} & \qw
        \end{quantikz}
        \caption{Grover's coin.}
        \label{subfig:grover}
    \end{subfigure}
     \caption{Coin operators used in the DTQW implementation.}
    \label{fig:coins}
\end{figure}

The coin operator determines how the amplitude of the walker is distributed among the possible movement directions before the shift is applied. For a one-dimensional discrete-time quantum walk, a Hadamard coin is commonly used to create an equal superposition between left and right movement. For higher-dimensional walks, Grover-type diffusion coins are often used because they distribute amplitude over multiple outgoing directions. Figure~\ref{fig:coins} illustrates the basic coin circuits used as building blocks in the quantum walk implementation. In the lackadaisical case, the coin space is enlarged by an additional self-loop state, and the corresponding coin operator becomes a weighted Grover-type reflection.

%A quantum circuit is the schematic representation of any quantum algorithm or quantum program. Each line in the quantum circuit is expressed as a qubit and the operations, i.e., quantum gates are illustrated by different blocks on the line. The gates used in this implementation are mentioned in the table ~\ref{tab:quantum_gates_matrices} with the matrix and graphical representation. 

%The generic circuits of various coins and shift operations used in discrete-time quantum walks are discussed in this part in two subsections. Firstly, the types of coins and their implementation will be discussed, and then the types of shift operations will be discussed. Here we are discussing two types of coins: the Hadamard coin and Grover's Coin. The Hadamard coin is the most-used two-dimensional coin, mimicking a fair coin by creating equal superposition on each state, and Grover's coin is used for higher dimensions; it distributes amplitude evenly across all available directional pathways. The circuital structures~\cite{qiskit-textbook} of the coins are given in Fig[~\ref{fig:coins}].

\begin{figure}[!h]
    \centering
    \begin{subfigure}[c]{0.48\textwidth}
        \centering
        \begin{quantikz}[column sep=0.4cm, row sep=0.35cm]
            \lstick{$|q_0\rangle$}   & \ctrl{2}  & \ctrl{1}  & \targ{}   & \qw \\
            \lstick{$|q_1\rangle$}   & \ctrl{1}  & \targ{}   & \qw       & \qw \\
            \lstick{$|q_2\rangle$}   & \targ{}   & \qw       & \qw       & \qw \\
            \lstick{$|coin\rangle$}  & \ctrl{-1} & \ctrl{-2} & \ctrl{-3} & \qw
        \end{quantikz}
        \caption{Increment Circuit $(S_+)$.}
        \label{subfig:increment}
    \end{subfigure}
    %\hfill
    \begin{subfigure}[c]{0.48\textwidth}
        \centering
        \begin{quantikz}[column sep=0.4cm, row sep=0.35cm]
            \lstick{$|q_0\rangle$}   & \qw       & \targ{}   & \ctrl{1}  & \ctrl{2}  & \qw       & \qw \\
            \lstick{$|q_1\rangle$}   & \qw       & \qw       & \targ{}   & \ctrl{1}  & \qw       & \qw \\
            \lstick{$|q_2\rangle$}   & \qw       & \qw       & \qw       & \targ{}   & \qw       & \qw \\
            \lstick{$|coin\rangle$}  & \gate{X}  & \ctrl{-3} & \ctrl{-2} & \ctrl{-1} & \gate{X}  & \qw
        \end{quantikz}
        \caption{Decrement Circuit $(S_-)$.}
        \label{subfig:decrement}
    \end{subfigure}
 
    \caption{Shift operator circuits for the discrete-time quantum walks
             on a 3-qubit position register. $S_+$ increments the
             position by 1 when $|coin\rangle = |1\rangle$;
             $S_-$ decrements by 1 when $|coin\rangle = |0\rangle$.}
    \label{fig:shift_circuits}
\end{figure}
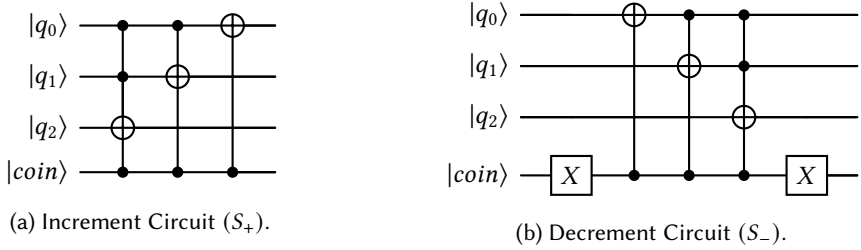

After the coin operation, a shift operator updates the position of the walker according to the coin state. In a one-dimensional DTQW, the shift can be implemented by controlled increment and decrement operations on the position register: one coin state increments the position, while the other decrements it \cite{Saha2024-fv}. These operations are reversible and can therefore be realized using controlled quantum gates. Figure~\ref{fig:shift_circuits} shows the increment and decrement circuits for a three-qubit position register. Although nearest-neighbor shift operators \cite{Singh_2021, 9410395} can also be defined directly at the operator level, the increment--decrement construction is more suitable for scalable register-based circuit implementation and is therefore used as the basis for the shift operations in this work.

%Two type of shift operations are discussed here: increment, decrement, ~\cite{saha2024robust}, and nearest neighbor ~\cite{singh2021quantum}. For increment-decrement, the walker's spatial position is updated based on the state of a ``coin" qubit. If the coin is in state \(\vert{}0\rangle\), the position decreases (decrement); if \(\vert{}1\rangle\), it increases (increment). The total shift operator (\(S\)) for a step is mathematically expressed as:\(S = S_{-} \otimes \vert{}0\rangle\langle0\vert{} + S_{+} \otimes \vert{}1\rangle\langle1\vert{}\).

%The nearest neighbor shift operator moves the quantum walker from a given lattice site to adjacent, neighboring sites depending on the state of the "coin" register. It creates the stepping action of the walks. In a one-dimensional walks, the shift operator is a unitary transformation that can be written as: 
\begin{comment}

\begin{center}
\(\hat{S} = \sum_{x} \left( \vert{}x+1\rangle\langle x\vert{} \otimes \vert{}0\rangle\langle 0\vert{} + \vert{}x-1\rangle\langle x\vert{} \otimes \vert{}1\rangle\langle 1\vert{} \right)\).
    
\end{center}
Since this shift operation is specific to the nearest-neighbor case and does not readily generalize to all scenarios considered in this work, we instead employ a naive increment-decrement operation, as illustrated previously.
\end{comment}

\begin{figure}[!h]
    \centering
    % --- (a) Oracle for |00> ---
    \begin{subfigure}[c]{0.32\textwidth}
        \centering
        \begin{quantikz}[column sep=0.4cm, row sep=0.55cm]
            \lstick{\ket{q_0}}  & \gate{X} & \ctrl{1}  & \gate{X} & \qw \\
            \lstick{\ket{q_1}}  & \gate{X} & \control{} & \gate{X} & \qw 
        \end{quantikz}
        \caption{State $\ket{00}$}
        \label{subfig:oracle_00}
    \end{subfigure}
    \hfill
    % --- (b) Oracle for |01> ---
    \begin{subfigure}[c]{0.32\textwidth}
        \centering
        \begin{quantikz}[column sep=0.4cm, row sep=0.55cm]
            \lstick{\ket{q_0}}  & \qw & \ctrl{1}  & \qw      & \qw \\
            \lstick{\ket{q_1}}  & \gate{X} & \control{} & \gate{X} & \qw 
        \end{quantikz}
        \caption{State $\ket{01}$}
        \label{subfig:oracle_01}
    \end{subfigure}
    \hfill
    % --- (c) Oracle for |10> ---
    \begin{subfigure}[c]{0.32\textwidth}
        \centering
        \begin{quantikz}[column sep=0.4cm, row sep=0.55cm]
            \lstick{\ket{q_0}}  & \gate{X} & \ctrl{1}  & \gate{X} & \qw \\
            \lstick{\ket{q_1}}  & \qw & \control{} & \qw      & \qw 
        \end{quantikz}
        \caption{State $\ket{10}$}
        \label{subfig:oracle_10}
    \end{subfigure}
    \hfill
    % --- (d) Oracle for |11> ---
    \begin{subfigure}[c]{0.32\textwidth}
        \centering
        \begin{quantikz}[column sep=0.4cm, row sep=0.55cm]
            \lstick{\ket{q_0}}  & \qw & \ctrl{1}  & \qw      & \qw \\
            \lstick{\ket{q_1}}  & \qw & \control{} & \qw      & \qw 
        \end{quantikz}
        \caption{State $\ket{11}$}
        \label{subfig:oracle_11}
    \end{subfigure}

    \caption{The four possible Grover's oracle operators for a 2-qubit system.}
    \label{fig:grovers_oracle}
\end{figure}
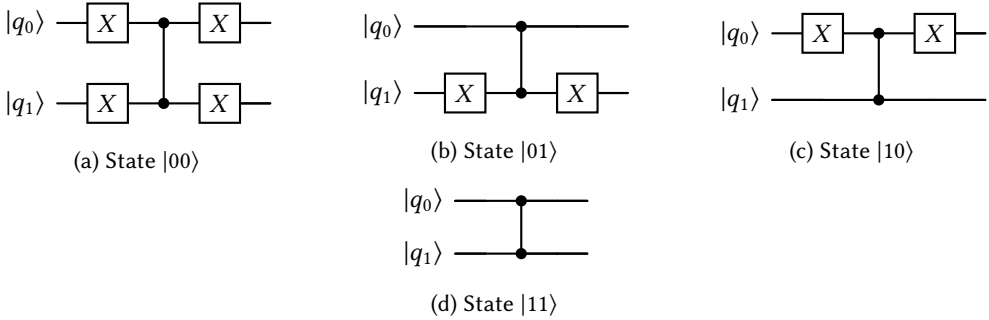

The final essential component is the phase oracle, which marks the target vertex in the search space. For a marked state, the oracle is defined as Eq. \ref{eq:phase_oracle}. This operation flips the phase of the marked state while leaving all other basis states unchanged. For a two-qubit search space, different marked states can be selected by surrounding a controlled-phase operation with $X$ gates on the qubits whose target bit value is $0$. Figure~\ref{fig:grovers_oracle} illustrates this construction for all four computational basis states. 

\begin{comment}

After the discussion of the shift operation, the remaining core component of the quantum walks framework is the Phase Oracle. The oracle utilized here is a standard Grover’s oracle, $G$, which is formally defined as:

\begin{center}
    $G = I_N - 2\ket{w}\bra{w}$
\end{center}

where $I_N$ denotes the identity operator for an $N$-dimensional space and $\ket{w}$ represents the target or ``marked'' state. Geometrically, this operator performs a selective phase inversion, flipping the sign of the amplitude of the marked state while leaving all other orthogonal basis states completely unaltered: $G\ket{w} = -\ket{w}$, and $G\ket{x} = \ket{x}$ for all $\ket{x} \neq \ket{w}$.

For a 2-qubit system ($N=4$), the search space is spanned by the computational basis states $\{\ket{00}, \ket{01}, \ket{10}, \ket{11}\}$. In a circuit implementation, this phase inversion is physically realized using a Controlled-Z ($CZ$) gate architecture. To target states containing a $\ket{0}$ bit, bit-flip ($X$) gates are applied as pre- and post-activation wrappers around the respective qubit wires. This temporarily transforms the $\ket{0}$ state into a $\ket{1}$ state to trigger the phase-flip condition, before restoring the original state of the register. The explicit quantum circuit representations for all four possible single-marked states in a 2-qubit system are systematically illustrated in Fig.~\ref{fig:grovers_oracle}.
\end{comment}

In the full lackadaisical quantum walk search circuit, this oracle acts on the position register, while the coin register is left unchanged.
Together, these circuit elements provide the background required for the proposed implementation. The Hadamard and Grover-type coins define the local mixing operation, the controlled increment and decrement circuits implement position updates, and the phase oracle introduces the marked state phase inversion required for search. Section~\ref{sec:impl} builds on these components to construct the complete gate-level circuit for lackadaisical quantum walk search on two-dimensional grids.

\section{Circuit design of lackadaisical quantum walk search}
\label{sec:impl}

This section presents the gate-level implementation of the lackadaisical quantum walk search algorithm on a two-dimensional grid. The objective is to translate the abstract walk operator into a quantum circuit consisting of position registers, a coin register, state preparation, a phase oracle, an embedded lackadaisical coin, and a controlled flip-flop shift operator. We first describe the register encoding and initialization procedure, then define the oracle, coin, and shift operators used in the circuit. Finally, we give the complete implementation procedure and show that the implemented circuit realizes the intended LQW search step on the valid computational subspace.

\subsection{Register encoding}
\label{subsec:encoding}

We consider a two-dimensional $L \times L$ grid with $N=L^2$ vertices. For circuit implementation, we assume $L=2^n$, so that each coordinate can be encoded using an $n$-qubit register. The position of the walker is represented by two registers,
\begin{equation}
X_p=(x_1,x_2,\ldots,x_n), \qquad
Y_p=(y_1,y_2,\ldots,y_n),
\end{equation}
where $X_p$ stores the horizontal coordinate and $Y_p$ stores the vertical coordinate. Therefore, the position Hilbert space is
\begin{equation}
\mathcal{H}_P=\mathcal{H}_X \otimes \mathcal{H}_Y,
\qquad
\dim(\mathcal{H}_P)=N.
\end{equation}

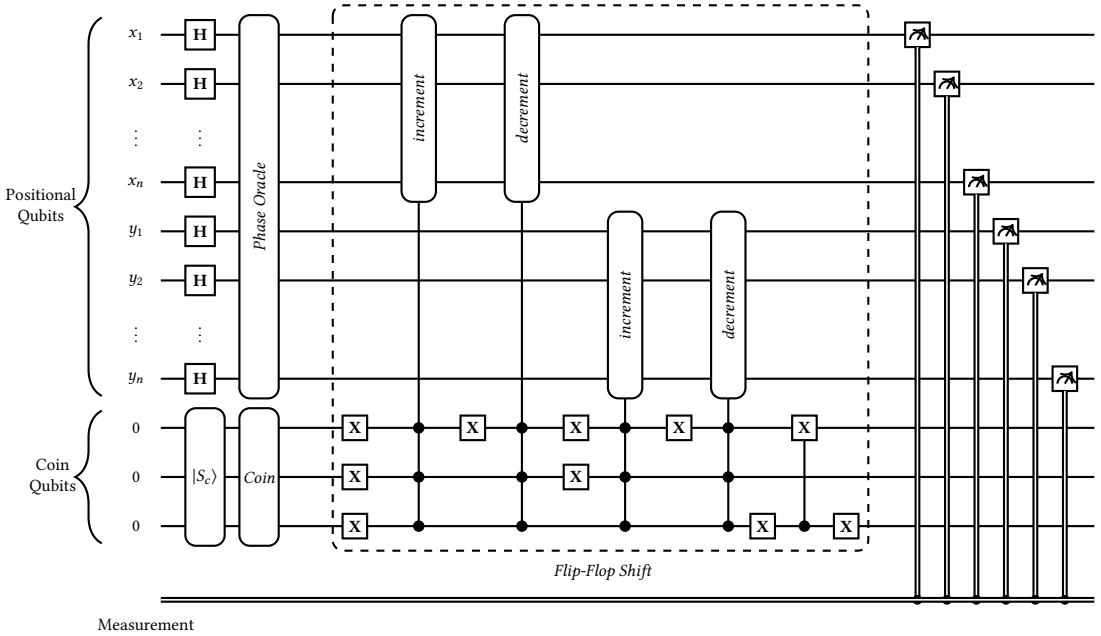
\begin{figure}[!h]
    \centering
\begin{tikzpicture}[scale=0.65, every node/.style={transform shape}]

% Define Wires with the new gaps
% x1=1, x2=2, (gap=3), xn=4, y1=5, y2=6, (gap=7), yn=8, Coin=9, Coin=10, Coin=11
\foreach \y / \label in {1/$x_1$, 2/$x_2$, 4/$x_n$, 5/$y_1$, 6/$y_2$, 8/$y_n$, 9/$0$, 10/$0$, 11/$0$} {
    \node at (-0.5, -\y) (L\y) {\label};
    \draw[thick] (0, -\y) -- (19.0, -\y);
}
\draw [thick, decorate, decoration={brace, amplitude=10pt, mirror}] 

      (-1.2, -0.65) -- (-1.2, -8.35);

\node[align=center, anchor=east] at (-1.6, -4.5) {Positional \\ Qubits};
\draw [thick, decorate, decoration={brace, amplitude=10pt, mirror}] 

      (-1.2, -8.65) -- (-1.2, -11.35);

\node[align=center, anchor=east] at (-1.6, -10.0) {Coin \\ Qubits};
% vdots for gaps in X and Y registers
\node at (-0.5, -3) {$\vdots$};
\node at (0.8, -3) {$\vdots$};
\node at (-0.5, -7) {$\vdots$};
\node at (0.8, -7) {$\vdots$};

% H Gates
\foreach \y in {1,2,4,5,6,8} {
    \draw[fill=white, thick] (0.5, -\y-0.3) rectangle (1.1, -\y+0.3);
    \node at (0.8, -\y) {\textbf{H}};
}

% Sc State block
\draw[fill=white, rounded corners, thick] (0.5, -11.4) rectangle (1.3, -8.6);
\node at (0.9, -10.0) {$|S_c\rangle$};

% Coin Block
\draw[fill=white, rounded corners, thick] (1.6, -11.4) rectangle (2.4, -8.6);
\node at (2.0, -10.0) {\textit{Coin}};

% X gates group 1
\foreach \y in {9,10,11} {
    \draw[fill=white, thick] (3.7, -\y-0.25) rectangle (4.2, -\y+0.25);
    \node at (3.95, -\y) {\textbf{X}};
}

% Increment X Box
\draw[fill=white, rounded corners, thick] (4.9, -4.4) rectangle (5.6, -0.6);
\node[rotate=90] at (5.25, -2.5) {\textit{increment}};
\draw[thick] (5.25, -4.4) -- (5.25, -11);
\foreach \y in {9,10,11} \filldraw (5.25, -\y) circle (3pt);

% X gates group 2
\draw[fill=white, thick] (6.1, -9.25) rectangle (6.6, -8.75); \node at (6.35, -9) {\textbf{X}};

% Decrement X Box
\draw[fill=white, rounded corners, thick] (7.0, -4.4) rectangle (7.7, -0.6);
\node[rotate=90] at (7.35, -2.5) {\textit{decrement}};
\draw[thick] (7.35, -4.4) -- (7.35, -11);
\foreach \y in {9,10,11} \filldraw (7.35, -\y) circle (3pt);

% X gates group 3
\draw[fill=white, thick] (8.2, -9.25) rectangle (8.7, -8.75); \node at (8.45, -9) {\textbf{X}};
\draw[fill=white, thick] (8.2, -10.25) rectangle (8.7, -9.75); \node at (8.45, -10) {\textbf{X}};

% Increment Y Box
\draw[fill=white, rounded corners, thick] (9.1, -8.4) rectangle (9.8, -4.6);
\node[rotate=90] at (9.45, -6.5) {\textit{increment}};
\draw[thick] (9.45, -8.4) -- (9.45, -11);
\foreach \y in {9,10,11} \filldraw (9.45, -\y) circle (3pt);

% X gates group 4
\draw[fill=white, thick] (10.3, -9.25) rectangle (10.8, -8.75); \node at (10.55, -9) {\textbf{X}};

% Decrement Y Box
\draw[fill=white, rounded corners, thick] (11.2, -8.4) rectangle (11.9, -4.6);
\node[rotate=90] at (11.55, -6.5) {\textit{decrement}};
\draw[thick] (11.55, -8.4) -- (11.55, -11);
\foreach \y in {9,10,11} \filldraw (11.55, -\y) circle (3pt);

% PhaseOracle Box
\draw[fill=white, rounded corners, thick] (1.6, -8.4) rectangle 
(2.4, -0.6);
\node[rotate=90] at (2, -4.5) {\textit{Phase Oracle}};

\draw[thick, dashed, rounded corners] (3.5, -0.4) rectangle 
(14.4,-11.5);
\node[below] at (9, -11.65) {\textit{Flip-Flop Shift}};

% First X gate on bottom Coin wire
\draw[fill=white, thick] (12, -11.25) rectangle (12.5, -10.75); \node at (12.25, -11) {\textbf{X}};

% Toffoli-like gate 
\draw[thick] (13.1, -9) -- (13.1, -11);
\filldraw (13.1, -11) circle (3pt);
\draw[fill=white, thick] (12.85, -9.25) rectangle (13.35, -8.75); \node at (13.1, -9) {\textbf{X}};

% Second X gate on bottom Coin wire
\draw[fill=white, thick] (13.7, -11.25) rectangle (14.2, -10.75); \node at (13.95, -11) {\textbf{X}};

\foreach \i/\y in {1/1, 2/2, 3/4, 4/5, 5/6, 6/8} {
    \pgfmathsetmacro{\xpos}{14.8 + \i * 0.6}
    \draw[fill=white, thick] (\xpos-0.25, -\y-0.25) rectangle (\xpos+0.25, -\y+0.25);
    \draw[thick] (\xpos-0.15, -\y-0.1) arc (180:0:0.15);
    \draw[->, thick] (\xpos, -\y-0.1) -- (\xpos+0.15, -\y+0.15);
    
    \draw[double, thick] (\xpos, -\y-0.25) -- (\xpos, -12.5);
    \filldraw (\xpos, -12.5) circle (2.5pt);
}

\node at (-0.3, -13) (M) {\text{Measurement}};
\draw[double, thick] (0, -12.5) -- (19.0, -12.5);

\end{tikzpicture}  
\caption{Complete circuit design of the lackadaisical quantum walks on an $L \times L$ grid, showing state initialization, the walk operator repeated for $t$ iterations (coin, phase oracle, and flip-flop shift operations), and final measurement of the vertex qubits.}
\label{fig:circuit}
\end{figure}

As discussed in Eq. \ref{eq:lqwcoinspace}, in the lackadaisical model, an additional self-loop state is introduced so that the walker can remain at its current vertex. Hence, the logical coin space is five-dimensional. Since a five-dimensional coin space cannot be represented directly using an integer number of qubits, we embed it into a three-qubit register. With this encoding, the complete circuit acts on $2n+3$ qubits. Figure~\ref{fig:circuit} shows the complete circuit design based on the exact mathematical walk step used in the implementation is
defined in Eq. \ref{eq:lqwsearch}.

\begin{comment}

The three-qubit coin register has eight computational basis states, of which five are used as valid coin states:
\begin{align}
\ket{\uparrow} &\equiv |000\rangle, &
\ket{\downarrow} &\equiv |001\rangle, &
\ket{\leftarrow} &\equiv |010\rangle, &
\ket{\rightarrow} &\equiv |011\rangle, &
\ket{\circlearrowleft} &\equiv |100\rangle.
\end{align}
The remaining states $|101\rangle$, $|110\rangle$, and $|111\rangle$ are unused auxiliary states. 
\end{comment} 

 The upper two groups of qubits are the position registers $X_p$ and $Y_p$, while the lower three qubits form the embedded coin register. After initialization, the coin block applies a lackadaisical coin. The controlled increment and decrement blocks implement the position update, which is part of the shift. The surrounding $X$ gates on the coin qubits convert each required coin condition into the all-one control pattern, so that each arithmetic block is activated only for its intended direction. The phase-oracle block applies the sign flip to the marked position state or states. The $CX$ gate is used for the direction-reversal operation on the coin register, which maps each moving direction to its opposite direction after the corresponding position update. Finally, the measurement symbols indicate that only the position registers are read out to obtain the output vertex distribution. Together, these modules implement one LQW search iteration; repeating the iteration $t$ times yields the final state before measurement.

\subsection{Initialization } 
\label{subsec:ini}

The initial state is a uniform superposition over all grid vertices together with the lackadaisical coin state as defined in Eq. \ref{eq:initvertex}. This state is prepared by applying Hadamard gates to all $2n$ position qubits.

The lackadaisical coin state depends on the self-loop weight $\ell$. It is defined as in Eq. \ref{eq:lqwcoin}. Under the three-qubit encoding, this state is represented by the eight-dimensional vector
\begin{equation}
|s_{C_8}^{(\ell)}\rangle
=
\begin{bmatrix}
\frac{1}{\sqrt{4+\ell}} \
\frac{1}{\sqrt{4+\ell}} \
\frac{1}{\sqrt{4+\ell}} \
\frac{1}{\sqrt{4+\ell}} \
\frac{\sqrt{\ell}}{\sqrt{4+\ell}} \
0 \
0 \
0
\end{bmatrix}.
\end{equation}
Therefore, the complete initial state of the walk is
\begin{equation}
|\Psi_0\rangle
=
|s_P\rangle \otimes |s_{C_8}^{(\ell)}\rangle .
\end{equation}

The decomposed circuit realizes the hardware-compatible initialization of the five-dimensional lackadaisical coin inside the eight-dimensional Hilbert
space of three qubits.

\subsection{Lackadaisical coin operator}
\label{subsec:coin}
The lackadaisical coin is a Grover-type reflection about the weighted coin state $|s_C^{(\ell)}\rangle$. On the logical five-dimensional coin space, the coin operator is defined as in Eq. \ref{eq:lqwfivedimcoin}. Since the coin is implemented using three qubits, the five-dimensional coin operator is embedded into an eight-dimensional unitary operator:
\begin{equation}
C_8^{(\ell)}
=
\begin{bmatrix}
C_5^{(\ell)} & 0 \
0 & I_3
\end{bmatrix},
\end{equation}
where $I_3$ acts on the unused auxiliary states $|101\rangle$, $|110\rangle$, and $|111\rangle$. This embedding preserves the desired lackadaisical coin operation on the valid five-dimensional coin subspace and leaves the unused states unchanged. In the full circuit, the coin operation is applied as
\begin{equation}
I_N \otimes C_8^{(\ell)}.
\end{equation}

Therefore, the definition of the proposed lackadaisical coin matrix is as follows.:
\begin{equation}
    \setlength{\arraycolsep}{4pt}
    C_8^{(\ell)} = \begin{bmatrix}
        \frac{2}{4+\ell} - 1     & \frac{2}{4+\ell}         & \frac{2}{4+\ell}         & \frac{2}{4+\ell}         & \frac{2\sqrt{\ell}}{4+\ell} & 0 & 0 & 0 \\
        \frac{2}{4+\ell}         & \frac{2}{4+\ell} - 1     & \frac{2}{4+\ell}         & \frac{2}{4+\ell}         & \frac{2\sqrt{\ell}}{4+\ell} & 0 & 0 & 0 \\
        \frac{2}{4+\ell}         & \frac{2}{4+\ell}         & \frac{2}{4+\ell} - 1     & \frac{2}{4+\ell}         & \frac{2\sqrt{\ell}}{4+\ell} & 0 & 0 & 0 \\
        \frac{2}{4+\ell}         & \frac{2}{4+\ell}         & \frac{2}{4+\ell}         & \frac{2}{4+\ell} - 1     & \frac{2\sqrt{\ell}}{4+\ell} & 0 & 0 & 0 \\
        \frac{2\sqrt{\ell}}{4+\ell} & \frac{2\sqrt{\ell}}{4+\ell} & \frac{2\sqrt{\ell}}{4+\ell} & \frac{2\sqrt{\ell}}{4+\ell} & \frac{2\ell}{4+\ell} - 1     & 0 & 0 & 0 \\
        0                     & 0                     & 0                     & 0                     & 0                     & 1 & 0 & 0 \\
        0                     & 0                     & 0                     & 0                     & 0                     & 0 & 1 & 0 \\
        0                     & 0                     & 0                     & 0                     & 0                     & 0 & 0 & 1
    \end{bmatrix}.
\end{equation}

%\noindent\textbf{Lemma 1.}
\begin{lemma}

The embedded lackadaisical coin $C_8^{(\ell)}$ is unitary for every $\ell \geq 0$.
\end{lemma}
%\noindent\textit{Proof.}
\begin{proof}

The state $|s_C^{(\ell)}\rangle$ is normalized by construction. Hence,
\begin{equation}
C_5^{(\ell)}
=
2|s_C^{(\ell)}\rangle\langle s_C^{(\ell)}|
-
I_5
\end{equation}
is a reflection operator. Therefore,
\begin{equation}
\left(C_5^{(\ell)}\right)^\dagger C_5^{(\ell)} = I_5.
\end{equation}
Since $C_8^{(\ell)}$ is the direct sum of $C_5^{(\ell)}$ and $I_3$, we have
\begin{equation}
\left(C_8^{(\ell)}\right)^\dagger C_8^{(\ell)}
=
\begin{bmatrix}
\left(C_5^{(\ell)}\right)^\dagger C_5^{(\ell)} & 0 \
0 & I_3
\end{bmatrix}
=
I_8.
\end{equation}
Thus, $C_8^{(\ell)}$ is unitary.
\end{proof}

%Appendix~\ref{app:coin_decomposition} shows 

The gate-level decomposition of the lackadaisical coin operator acts exclusively on the three coin qubits and implements the coin-flip operation applied at each step of the lackadaisical quantum walk search. The single-qubit gates perform the required local basis transformations and phase adjustments, while the $CX$ gates introduce the necessary entangling operations among the coin qubits. Although the controlled increment and decrement operations update the walker’s position on the spatial lattice, they do not by themselves complete the walk-step transformation. To maintain the correct quantum walk dynamics and enable subsequent interference, the coin state must also be updated to encode the direction opposite to the performed displacement. This direction reversal is implemented by the flip-flop shift operation, which maps each movement coin state to its corresponding opposite direction after the position shift, which is discussed next.

%Although such controlled operations effectively move the walker through the spatial lattice, the step transformation is still not complete. For further quantum interference, the internal state of the coin should be transformed right away to take account of the new direction, and for that, the flip-flop procedure is required.

\subsection{Flip-flop shift operator}
\label{subsec:flipflop}

The flip-flop shift \cite{ambainis2004coins} moves the walker according to the active coin direction and then replaces the coin direction by its opposite. On the logical valid subspace, the shift is
\begin{equation}
S_{ff} : \begin{cases}
|x, y\rangle  \otimes  |\rightarrow\rangle  \rightarrow |(x + 1)\bmod L, y\rangle \otimes |\leftarrow\rangle  \\
|x, y\rangle  \otimes  |\leftarrow\rangle \rightarrow    |(x - 1)\bmod L, y\rangle \otimes |\rightarrow\rangle \\
  |x, y\rangle \otimes |\uparrow\rangle \rightarrow |x, (y + 1)\bmod L\rangle \otimes |\downarrow\rangle   \\
|x, y\rangle \otimes  |\downarrow\rangle  \rightarrow |x, (y - 1)\bmod L\rangle \otimes |\uparrow\rangle   \\
|x, y\rangle \otimes |\circlearrowright\rangle  \rightarrow |x, y\rangle \otimes |\circlearrowright\rangle. 
\end{cases}
\label{eq:sff}
\end{equation}
Thus the self-loop component is stationary, while each moving component is shifted and then relabeled with the opposite direction.

In the three-qubit implementation, the shift is realized as a reversible extension $\widetilde{S}_{\rm ff}$ of $S_{\rm ff}^{(5)}$. Controlled increment and decrement circuits act on the $X_p$ and $Y_p$ position registers, and $X$ gates on the coin register temporarily transform the desired active coin state into $\ket{111}$ so that standard multi-controlled gates can be used. The coin is then restored to the logical basis and the flip-flop relabeling is applied using the $CX$ and $X$ gates. On the valid coin states, this relabeling is
\begin{equation}
    \ket{000}\leftrightarrow\ket{001},
    \qquad
    \ket{010}\leftrightarrow\ket{011},
    \qquad
    \ket{100}\mapsto\ket{100}.
    \label{eq:coin_flipflop_binary}
\end{equation}
Equivalently, on the valid subspace, the least significant coin bit is toggled exactly for moving states and not for the self-loop state.

\begin{lemma}

The implemented shift $\widetilde{S}_{\rm ff}$ is unitary. Moreover, its restriction to the valid coin subspace is the logical flip-flop shift $S_{\rm ff}$ in Eq.~\ref{eq:sff}.
\end{lemma}
\begin{proof}

Each gate used in the shift circuit is reversible. Therefore the total circuit implements a permutation of the computational basis states of the full position--coin Hilbert space, and every such permutation is unitary. For the five valid coin states, the controlled increment/decrement operations and the final relabeling in Eq.~\ref{eq:coin_flipflop_binary} give exactly the five mappings in Eq.~\ref{eq:sff}. Hence the implemented shift is a unitary reversible extension of the logical flip-flop shift.
\end{proof}

%Figure~\ref{fig:ffs} shows the gate-level flip-flop shift circuit for three-qubit $X_p$ and $Y_p$ coordinate registers, corresponding to an $8\times 8$ grid as an example. The circuit uses two three-qubit position registers, $X_p=(x_1,x_2,x_3)$ and $Y_p=(y_1,y_2,y_3)$, to encode the horizontal and vertical coordinates, together with a three-qubit coin register $(c_1,c_2,c_3)$. The five valid coin states encode motion along the four lattice directions and the self-loop state.

\begin{figure}[!h]
    \centering
    \begin{tikzpicture}[scale=0.72, every node/.style={transform shape}]
 
        % ── QUANTUM WIRES 
        \node at (-0.8, -1) {$x_1$};  \draw[thick] (0, -1) -- (17.5, -1);
        \node at (-0.8, -2) {$x_2$};  \draw[thick] (0, -2) -- (17.5, -2);
        \node at (-0.8, -3) {$x_3$};  \draw[thick] (0, -3) -- (17.5, -3);
        \node at (-0.8, -4) {$y_1$};  \draw[thick] (0, -4) -- (17.5, -4);
        \node at (-0.8, -5) {$y_2$};  \draw[thick] (0, -5) -- (17.5, -5);
        \node at (-0.8, -6) {$y_3$};  \draw[thick] (0, -6) -- (17.5, -6);
        \node at (-0.8, -7) {$c_1$}; \draw[thick] (0, -7) -- (17.5, -7);
        \node at (-0.8, -8) {$c_2$}; \draw[thick] (0, -8) -- (17.5, -8);
        \node at (-0.8, -9) {$c_3$}; \draw[thick] (0, -9) -- (17.5, -9);
 
        % ── STATE BOUNDARY LINES (12 LINES MATCHING THE GREEN STRIPES)
        % Drawn BEFORE gates so that white-filled gate boxes overlay them cleanly.
        \foreach \xpos/\idx in {%
            0.08/0,%   1: Before Coin block
            1.35/1,%   2: After Coin block, before initial X gates
           % 2.12/2,%   3: After initial X gates
            4.52/2,%   4: Inside x-shift, before first X-inversions
            %5.27/4,%   5: After mid-stage flip on Coin1
            7.62/3,%   6: After x-shift completes, before intermediate coins
            %8.47/6,%   7: After intermediate coins, before y-shift
            10.9/4,%  8: Inside y-shift stage
            14.07/5,%  9: After y-shift completes, before cleanup
            15.02/6,%  10: Inside final cleanup sequence
            16.07/7,% 11: Inside final cleanup sequence
            17.20/8%  12: After final cleanup sequence
        }{%
            \draw[dashed, thick, blue] (\xpos, -0.6) -- (\xpos, -9.6);
            \node[below] at (\xpos, -9.6) {$\lvert\psi_{\idx}\rangle$};
        }
 
        % ── COIN OPERATOR INITIAL BLOCK 
        \draw[fill=white, rounded corners=4pt, thick] (0.2, -9.3) rectangle (1.2, -6.7);
        \node at (0.7, -8) {$Coin$};
 
        % ── INITIAL X GATES ON ALL THREE COIN WIRES 
        \foreach \y in {-7, -8, -9} {
            \draw[fill=white, thick] (1.5, \y-0.25) rectangle (2.0, \y+0.25);
            \node at (1.75, \y) {\textbf{X}};
        }
 
        % ── X-DIRECTION SHIFT OPERATIONS 
 
        % Step 1: multi-controlled X on x₃
        \draw[thick] (2.5, -1) -- (2.5, -9);
        \filldraw (2.5, -1) circle (2.5pt);
        \filldraw (2.5, -2) circle (2.5pt);
        \filldraw (2.5, -7) circle (2.5pt);
        \filldraw (2.5, -8) circle (2.5pt);
        \filldraw (2.5, -9) circle (2.5pt);
        \draw[fill=white, thick] (2.25, -3.25) rectangle (2.75, -2.75);
        \node at (2.5, -3) {\textbf{X}};
 
        % Step 2: multi-controlled X on x₂
        \draw[thick] (3.3, -1) -- (3.3, -9);
        \filldraw (3.3, -1) circle (2.5pt);
        \filldraw (3.3, -7) circle (2.5pt);
        \filldraw (3.3, -8) circle (2.5pt);
        \filldraw (3.3, -9) circle (2.5pt);
        \draw[fill=white, thick] (3.05, -2.25) rectangle (3.55, -1.75);
        \node at (3.3, -2) {\textbf{X}};
 
        % Step 3: multi-controlled X on x₁
        \draw[thick] (4.1, -1) -- (4.1, -9);
        \filldraw (4.1, -7) circle (2.5pt);
        \filldraw (4.1, -8) circle (2.5pt);
        \filldraw (4.1, -9) circle (2.5pt);
        \draw[fill=white, thick] (3.85, -1.25) rectangle (4.35, -0.75);
        \node at (4.1, -1) {\textbf{X}};
 
        % Mid-point (x-stage): flip Coin₁
        \draw[fill=white, thick] (4.7, -7.25) rectangle (5.2, -6.75);
        \node at (4.9, -7) {\textbf{X}};
 
        % Step 5: multi-controlled X on x₁ (second pass)
        \draw[thick] (5.6, -1) -- (5.6, -9);
        \filldraw (5.6, -7) circle (2.5pt);
        \filldraw (5.6, -8) circle (2.5pt);
        \filldraw (5.6, -9) circle (2.5pt);
        \draw[fill=white, thick] (5.35, -1.25) rectangle (5.85, -0.75);
        \node at (5.6, -1) {\textbf{X}};
 
        % Step 6: multi-controlled X on x₂ (second pass)
        \draw[thick] (6.4, -1) -- (6.4, -9);
        \filldraw (6.4, -1) circle (2.5pt);
        \filldraw (6.4, -7) circle (2.5pt);
        \filldraw (6.4, -8) circle (2.5pt);
        \filldraw (6.4, -9) circle (2.5pt);
        \draw[fill=white, thick] (6.15, -2.25) rectangle (6.65, -1.75);
        \node at (6.4, -2) {\textbf{X}};
 
        % Step 7: multi-controlled X on x₃ (second pass)
        \draw[thick] (7.2, -1) -- (7.2, -9);
        \filldraw (7.2, -1) circle (2.5pt);
        \filldraw (7.2, -2) circle (2.5pt);
        \filldraw (7.2, -7) circle (2.5pt);
        \filldraw (7.2, -8) circle (2.5pt);
        \filldraw (7.2, -9) circle (2.5pt);
        \draw[fill=white, thick] (6.95, -3.25) rectangle (7.45, -2.75);
        \node at (7.2, -3) {\textbf{X}};
 
        % ── INTERMEDIATE COIN FLIPS (Coin₁ and Coin₂) 
        \draw[fill=white, thick] (7.8, -7.25) rectangle (8.3, -6.75);
        \node at (8.0, -7) {\textbf{X}};
 
        \draw[fill=white, thick] (7.8, -8.25) rectangle (8.3, -7.75);
        \node at (8.0, -8) {\textbf{X}};
 
        % ── Y-DIRECTION SHIFT OPERATIONS 
 
        % Step 1: multi-controlled X on y₃
        \draw[thick] (8.9, -4) -- (8.9, -9);
        \filldraw (8.9, -4) circle (2.5pt);
        \filldraw (8.9, -5) circle (2.5pt);
        \filldraw (8.9, -7) circle (2.5pt);
        \filldraw (8.9, -8) circle (2.5pt);
        \filldraw (8.9, -9) circle (2.5pt);
        \draw[fill=white, thick] (8.65, -6.25) rectangle (9.15, -5.75);
        \node at (8.9, -6) {\textbf{X}};
 
        % Step 2: multi-controlled X on y₂
        \draw[thick] (9.7, -4) -- (9.7, -9);
        \filldraw (9.7, -4) circle (2.5pt);
        \filldraw (9.7, -7) circle (2.5pt);
        \filldraw (9.7, -8) circle (2.5pt);
        \filldraw (9.7, -9) circle (2.5pt);
        \draw[fill=white, thick] (9.45, -5.25) rectangle (9.95, -4.75);
        \node at (9.7, -5) {\textbf{X}};
 
        % Step 3: multi-controlled X on y₁
        \draw[thick] (10.5, -4) -- (10.5, -9);
      %  \filldraw (10.5, -5) circle (2.5pt);
        \filldraw (10.5, -7) circle (2.5pt);
        \filldraw (10.5, -8) circle (2.5pt);
        \filldraw (10.5, -9) circle (2.5pt);
        \draw[fill=white, thick] (10.25, -4.25) rectangle (10.75, -3.75);
        \node at (10.5, -4) {\textbf{X}};
 
        % Mid-point (y-stage): flip Coin₁
        \draw[fill=white, thick] (11.1, -7.25) rectangle (11.6, -6.75);
        \node at (11.3, -7) {\textbf{X}};
 
        % Step 5: multi-controlled X on y₁ (second pass)
        \draw[thick] (12.0, -4) -- (12.0, -9);
        \filldraw (12.0, -7) circle (2.5pt);
        \filldraw (12.0, -8) circle (2.5pt);
        \filldraw (12.0, -9) circle (2.5pt);
        \draw[fill=white, thick] (11.75, -4.25) rectangle (12.25, -3.75);
        \node at (12.0, -4) {\textbf{X}};
 
        % Step 6: multi-controlled X on y₂ (second pass)
        \draw[thick] (12.8, -4) -- (12.8, -9);
        \filldraw (12.8, -4) circle (2.5pt);
        \filldraw (12.8, -7) circle (2.5pt);
        \filldraw (12.8, -8) circle (2.5pt);
        \filldraw (12.8, -9) circle (2.5pt);
        \draw[fill=white, thick] (12.55, -5.25) rectangle (13.05, -4.75);
        \node at (12.8, -5) {\textbf{X}};
 
        % Step 7: multi-controlled X on y₃ (second pass)
        \draw[thick] (13.6, -4) -- (13.6, -9);
        \filldraw (13.6, -4) circle (2.5pt);
        \filldraw (13.6, -5) circle (2.5pt);
        \filldraw (13.6, -7) circle (2.5pt);
        \filldraw (13.6, -8) circle (2.5pt);
        \filldraw (13.6, -9) circle (2.5pt);
        \draw[fill=white, thick] (13.35, -6.25) rectangle (13.85, -5.75);
        \node at (13.6, -6) {\textbf{X}};
 
        % ── FINAL COIN CLEANUP OPERATORS 
 
        % X on Coin₃
        \draw[fill=white, thick] (14.3, -9.25) rectangle (14.8, -8.75);
        \node at (14.5, -9) {\textbf{X}};
 
        % CNOT: Coin₁ (control) → Coin₃ (target)
        \draw[thick] (15.5, -7) -- (15.5, -9);
        \filldraw (15.5, -7) circle (2.5pt);
        \draw[fill=white, thick] (15.25, -9.25) rectangle (15.75, -8.75);
        \node at (15.5, -9) {\textbf{X}};
 
        % Final X on Coin₃
        \draw[fill=white, thick] (16.4, -9.25) rectangle (16.9, -8.75);
        \node at (16.65, -9) {\textbf{X}};
    \end{tikzpicture}
    \caption{Gate-level implementation of the flip-flop shift operator for an $8\times 8$ lackadaisical quantum walk. The circuit uses three-qubit $X_p$ and $Y_p$ position registers and a three-qubit coin register. Directional coin states are sequentially mapped to the active state $\ket{111}_c$ to control the corresponding spatial shift, followed by a flip-flop update that reverses the movement direction while preserving the self-loop state.}
    \label{fig:ffs}
\end{figure}
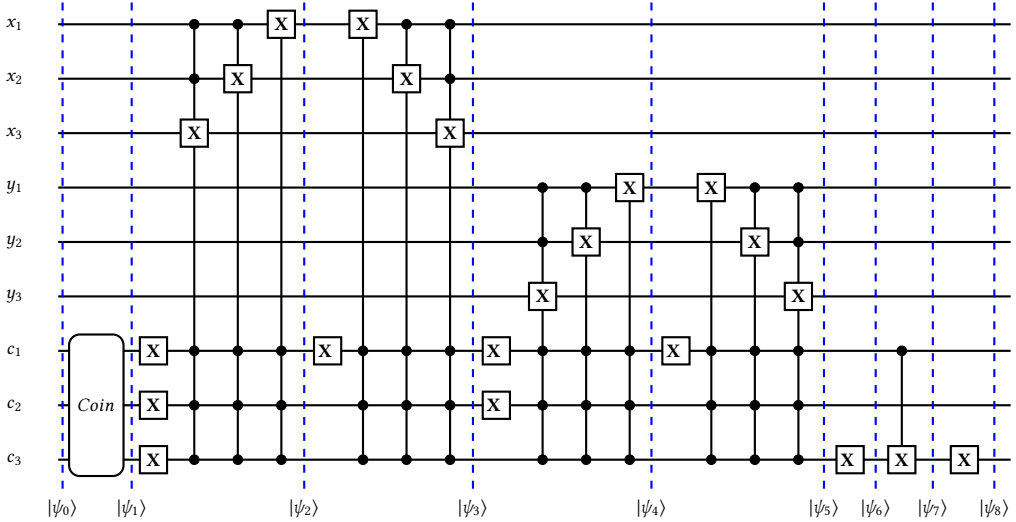

%Table~\ref{tab:flipflop_shift} traces the coin-state activations and the corresponding traversed vertices for the initial vertex $\ket{000,000}$.
Figure~\ref{fig:ffs} shows the gate-level implementation of the flip-flop
shift operator for the lackadaisical quantum walk on an $8\times 8$ grid.
We denote the two position registers by
\begin{equation}
    X_p=(x_1,x_2,x_3), \qquad
    Y_p=(y_1,y_2,y_3),
\end{equation}
where $X_p$ and $Y_p$ encode the horizontal and vertical coordinates of the
walker, respectively. The last three qubits form the coin register
$(c_1,c_2,c_3)$. The five valid lackadaisical coin states are encoded as
\begin{equation}
\ket{000}_c \equiv +X_p,\quad
\ket{001}_c \equiv -X_p,\quad
\ket{010}_c \equiv +Y_p,\quad
\ket{011}_c \equiv -Y_p,\quad
\ket{100}_c \equiv \circlearrowleft .
\end{equation}
Here, $\ket{100}_c$ denotes the self-loop state, for which no position
update is applied.

The shift is implemented by sequentially activating each directional coin
state. Since the multi-controlled shift blocks are triggered when the coin
register is in the state $\ket{111}_c$, $X$ gates are used to
temporarily map the required directional coin state to $\ket{111}_c$.
The corresponding controlled increment or decrement operation is then
applied to either the $X_p$ or $Y_p$ position register. After all directional
components have been shifted, the auxiliary coin transformations are
uncomputed, and the flip-flop update is applied. This final update maps each
movement direction to its opposite direction,
\begin{equation}
    +X_p \leftrightarrow -X_p,
    \qquad
    +Y_p \leftrightarrow -Y_p,
\end{equation}
while leaving the self-loop state unchanged.

The state evolution marked by the dashed boundaries
$\ket{\psi_0},\ket{\psi_1},\ldots,\ket{\psi_8}$ in
Fig.~\ref{fig:ffs} is described as follows:
\begin{itemize}
    \item \textbf{$\ket{\psi_0}\to\ket{\psi_1}$:}
    The lackadaisical coin operator is applied to the last three qubits.
    This prepares the coin register in a superposition of the four movement
    directions and the self-loop state.

    \item \textbf{$\ket{\psi_1}\to\ket{\psi_2}$:}
    The gates $X_{c_1}$, $X_{c_2}$, and $X_{c_3}$ map the $+X_p$ coin
    component to the active control state $\ket{111}_c$. The controlled
    shift then increments the $X_p$ register by one lattice unit. For the
    initial position $\ket{000,000}_{x,y}$, this gives
    \begin{equation}
        \ket{000,000}_{x,y}
        \longrightarrow
        \ket{001,000}_{x,y}.
    \end{equation}

    \item \textbf{$\ket{\psi_2}\to\ket{\psi_3}$:}
    The gate $X_{c_3}$ changes the active coin configuration so that the
    $-X_p$ component is mapped to $\ket{111}_c$. The corresponding
    controlled shift decrements the $X_p$ register by one lattice unit,
    producing
    \begin{equation}
        \ket{000,000}_{x,y}
        \longrightarrow
        \ket{111,000}_{x,y},
    \end{equation}
    according to the coordinate convention used in the circuit.

    \item \textbf{$\ket{\psi_3}\to\ket{\psi_4}$:}
    The gates $X_{c_2}$ and $X_{c_3}$ activate the $+Y_p$ component by
    mapping it to $\ket{111}_c$. The controlled shift increments the
    $Y_p$ register by one lattice unit, giving
    \begin{equation}
        \ket{000,000}_{x,y}
        \longrightarrow
        \ket{000,001}_{x,y}.
    \end{equation}

    \item \textbf{$\ket{\psi_4}\to\ket{\psi_5}$:}
    The gate $X_{c_3}$ activates the $-Y_p$ component. The corresponding
    controlled shift decrements the $Y_p$ register by one lattice unit,
    producing
    \begin{equation}
        \ket{000,000}_{x,y}
        \longrightarrow
        \ket{000,111}_{x,y},
    \end{equation}
    according to the coordinate convention used in the circuit. The
    self-loop component does not activate any controlled shift and therefore
    leaves the position state unchanged.

    \item \textbf{$\ket{\psi_5}\to\ket{\psi_6}$:}
    The auxiliary $X$ transformations used for directional activation
    are uncomputed. This restores the coin register to its logical basis
    before the flip-flop update is applied.

    \item \textbf{$\ket{\psi_6}\to\ket{\psi_7}$:}
    The anti-controlled flip-flop logic is initiated. This prepares the
    circuit to update only the movement coin states, while excluding the
    self-loop state from the direction reversal.

    \item \textbf{$\ket{\psi_7}\to\ket{\psi_8}$:}
    The controlled update is completed, implementing the flip-flop mapping
    \begin{equation}
        +X_p \leftrightarrow -X_p,
        \qquad
        +Y_p \leftrightarrow -Y_p,
    \end{equation}
    while preserving the self-loop state. Hence, the complete operator first
    moves the walker according to the active coin direction and then updates
    the coin label to the opposite direction, as required by the flip-flop
    shift operator.
\end{itemize}

\subsection{Phase oracle}
\label{subsec:oracle}

The oracle acts only on the position registers and flips the phase of the marked vertex or  vertices while leaving all unmarked vertices unchanged as described in Eq. \ref{eq:multiplemarked}. In the full position--coin Hilbert space, the oracle is applied as
\begin{equation}
Q_M \otimes I_8,
\end{equation}
where $I_8$ is the identity operator on the three-qubit coin register. In the circuit implementation, marked basis states are selected using multi-controlled phase operations. If a marked state contains zero-valued bits, $X$ gates are applied before and after the controlled phase operation to activate the desired computational basis state.

\subsection{Complete walk operator}
\label{subsec:complete_operator}

One complete step of the implemented lackadaisical quantum walk search circuit consists of the phase oracle, the embedded lackadaisical coin, and the flip-flop shift. From Eq. \ref{eq:lqwsearch}, the implemented walk operator can be defined as
\begin{equation}
U_{\mathrm{LQW}}^{(\ell)}
=
S_{\mathrm{ff}}
\left(I_N \otimes C_8^{(\ell)}\right)
\left(Q_M \otimes I_8\right).
\end{equation}
After $t$ walk steps, the state of the system becomes
\begin{equation}
|\Psi_t\rangle
=
\left(U_{\mathrm{LQW}}^{(\ell)}\right)^t
|\Psi_0\rangle.
\end{equation}
The position registers are then measured over $R$ shots. The success probability of observing a marked vertex is
\begin{equation}
p_M(t)
=
\sum_{w\in M}
\sum_{c}
\left|
\langle w,c|\Psi_t\rangle
\right|^2 .
\end{equation}

%\noindent\textbf{Proposition 1.}
\begin{proposition}

Restricted to the valid five-dimensional coin subspace, one iteration of the implemented circuit realizes the intended lackadaisical quantum walk search operator on the two-dimensional grid.
\end{proposition}
%\noindent\textit{Proof.}
\begin{proof}

The oracle $Q_M\otimes I_8$ applies the required phase inversion to the marked position states and acts trivially on the coin register. The embedded coin $I_N\otimes C_8^{(\ell)}$ applies the lackadaisical Grover reflection $C_5^{(\ell)}$ on the valid coin subspace and leaves the unused auxiliary states unchanged. The flip-flop shift $S_{\mathrm{ff}}$ then updates the position according to the active coin direction, while the self-loop state leaves the position unchanged.

The initial coin state has support only on the valid five-dimensional coin subspace. The embedded coin preserves this subspace, and the flip-flop shift maps valid direction states to valid direction states. Hence, no amplitude is introduced into the unused auxiliary coin states during the walk. Therefore, on the valid coin subspace, the implemented circuit applies
\begin{equation}
S_{\mathrm{ff}}
\left(I_N\otimes C_5^{(\ell)}\right)
\left(Q_M\otimes I_5\right),
\end{equation}
which is the intended lackadaisical quantum walk search step.

\end{proof}

Algorithm~\ref{alg:lqw_search} summarizes the gate-level construction of the lackadaisical quantum walk search circuit.

\begin{algorithm}[!h]
\caption{Gate-level implementation of lackadaisical quantum walk search}
\label{alg:lqw_search}
\begin{algorithmic}[1]
\State \textbf{Input:} Grid size $L=2^n$, marked set $M$, self-loop weight $\ell$, number of walk steps $t$, number of shots $R$
\State \textbf{Output:} Estimated marked-state success probability $p_M(t)$
\State Allocate two $n$-qubit position registers $X_p$ and $Y_p$
\State Allocate one three-qubit coin register $C$
\State Apply Hadamard gates to all qubits in $X_p$ and $Y_p$
\State Prepare $C$ in the state $|s_C^{(\ell)}\rangle$
\For{$j=1$ to $t$}
\State Apply the phase oracle $Q_M$ to the position registers
\State Apply the embedded lackadaisical coin $C_8^{(\ell)}$ to the coin register
\State Apply the controlled flip-flop shift $S_{\mathrm{ff}}$ to the position and coin registers
\EndFor
\State Measure the position registers over $R$ shots
\State Estimate $p_M(t)$ from the fraction of outcomes belonging to $M$
\State \Return $p_M(t)$
\end{algorithmic}
\end{algorithm}

This completes the circuit construction of the lackadaisical quantum walk search algorithm on a two-dimensional grid. To evaluate the effectiveness of the proposed design and assess its spatial-search performance, we next present a numerical analysis of the circuit under both ideal noiseless conditions and realistic operational constraints.

\section{Simulation results}\label{NR}
This section evaluates the proposed gate-level implementation of lackadaisical quantum walk search. The objective is twofold. First objective is to verify its correctness. We verify that the constructed circuit reproduces the expected search behavior in the ideal noiseless setting. Second, we study the effect of realistic noise, noise mitigation, and variation of the self-loop weight $\ell$ on the observed success probability.

From now on all experiments are performed on two-dimensional $L\times L$ grids, where $N=L^2$ is the total number of vertices. The position register therefore contains $\log_2 N$ qubits, and the lackadaisical coin is encoded using three additional qubits. The walker is initialized in a uniform superposition over all grid vertices, and the lackadaisical coin is initialized with the self-loop weight $\ell=4/N$. The circuit is then evolved for the prescribed number of walk steps, and the position register is measured to estimate the success probability of the marked vertex or multiple marked vertices. The implementation of LQW is available at \url{github.com/NishankaDas/Qwalks}.
\subsection{Ideal Noiseless Simulation}
We first evaluate the circuit in the absence of noise. This setting is used to confirm that the proposed gate-level construction correctly emulates the intended lackadaisical quantum walk search dynamics.
\subsubsection{Single marked state }

For the single marked case, we simulate grids of sizes $8\times 8$, $16\times 16$, $32\times 32$, and $64\times 64$, corresponding to $64$, $256$, $1024$, and $4096$ vertices, respectively. In each case, the marked vertex is chosen as a computational basis state of the form $|10\ldots 0\rangle$, so that the target is not concentrated near the all-zero initial computational basis state. The circuit is executed for the number of steps predicted by the lackadaisical quantum walk search procedure, and the measurement distribution over the position register is recorded. 

\begin{figure}[htpb]
    \centering
    \begin{subfigure}[b]{0.42\linewidth}
        \centering
        \includegraphics[width=\linewidth]{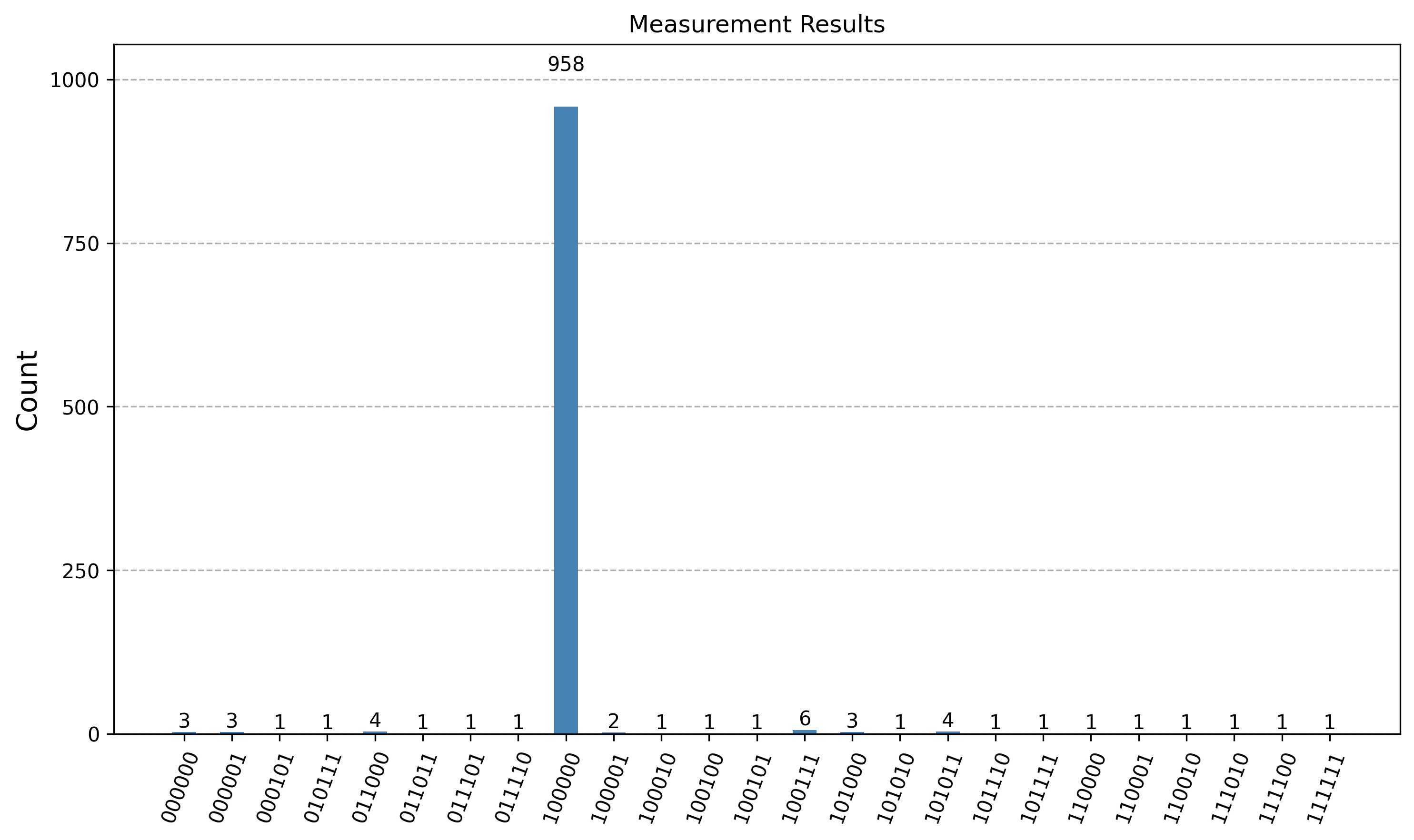}
        \caption{Final measurement distribution after $\sqrt{N\log_2 N}$ walk steps.}
        \label{fig:8x8_hist}
    \end{subfigure}
    \hfill
    \begin{subfigure}[b]{0.42\linewidth}
        \centering
        \includegraphics[width=\linewidth]{"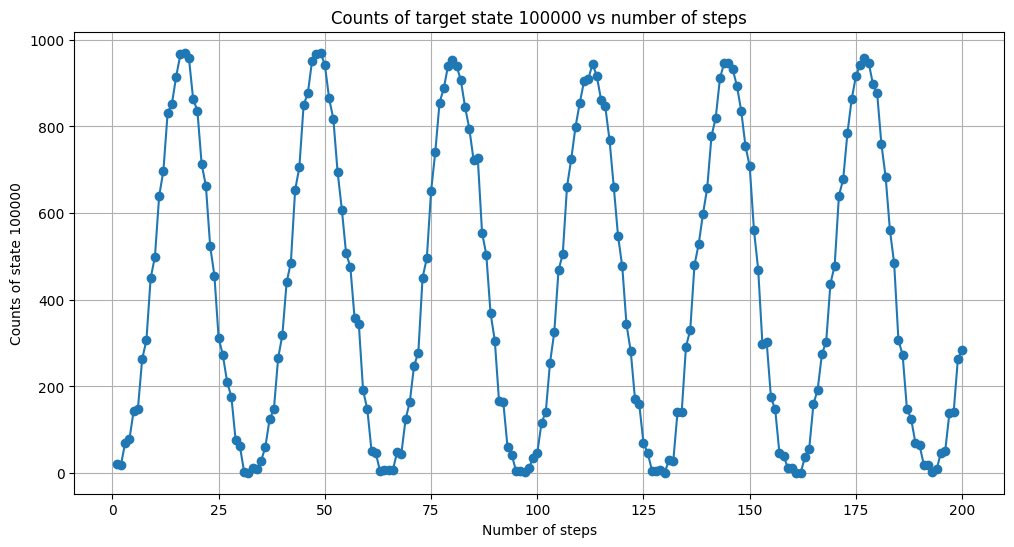"} 
        \caption{Marked state probability over walk steps.}
        \label{fig:8x8_prob}
    \end{subfigure}
    \caption{Ideal noiseless results for an $8\times 8$ grid with a single marked vertex.}
    \label{fig:grid_8x8}
\end{figure}

\begin{figure}[htpb]
    \centering
    \begin{subfigure}[b]{0.42\linewidth}
        \centering
        \includegraphics[width=\linewidth]{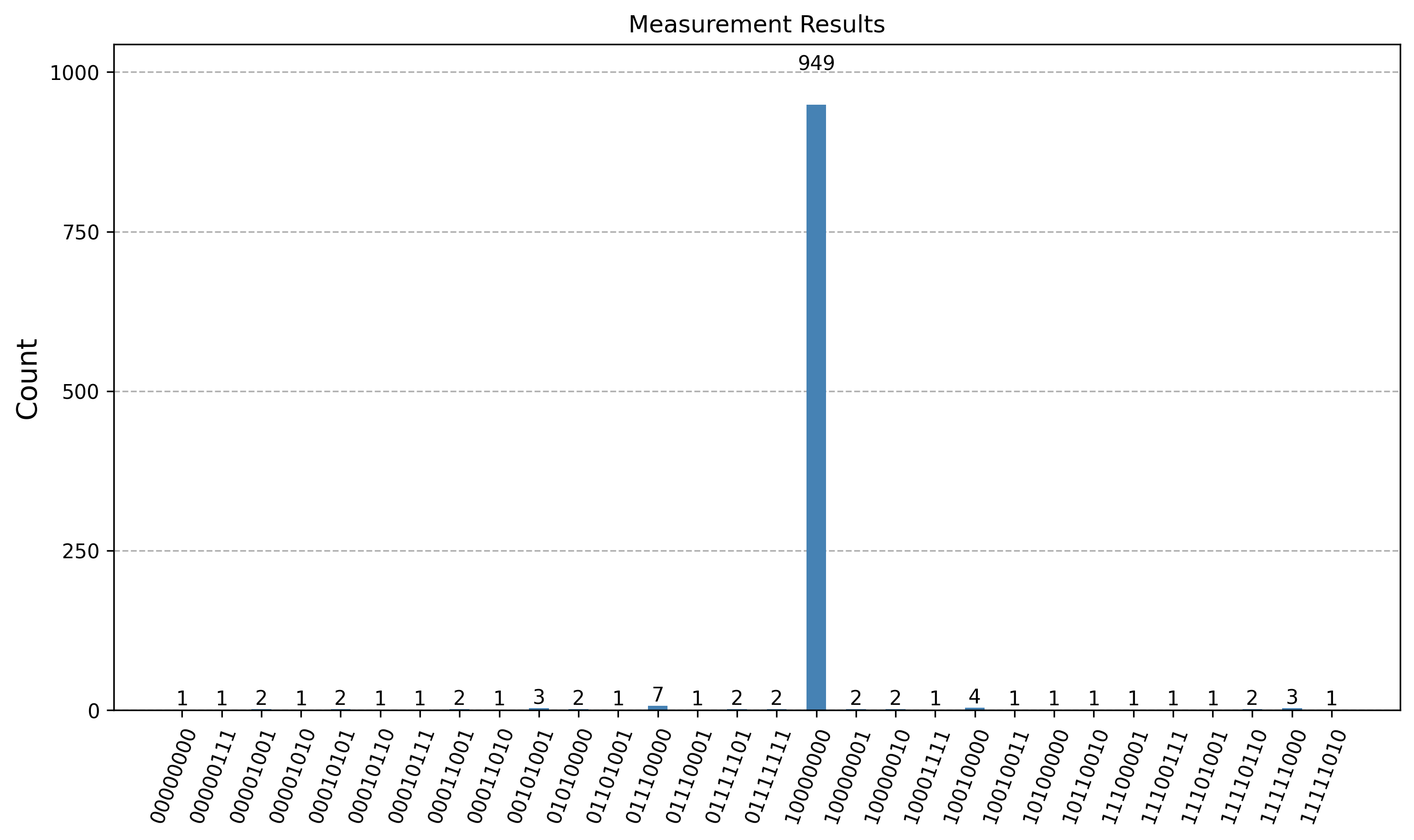}
        \caption{Final measurement distribution after $\sqrt{N\log_2 N}$ walk steps.}
        \label{fig:16x16_hist}
    \end{subfigure}
    \hfill
    \begin{subfigure}[b]{0.42\linewidth}
        \centering
        \includegraphics[width=\linewidth]{"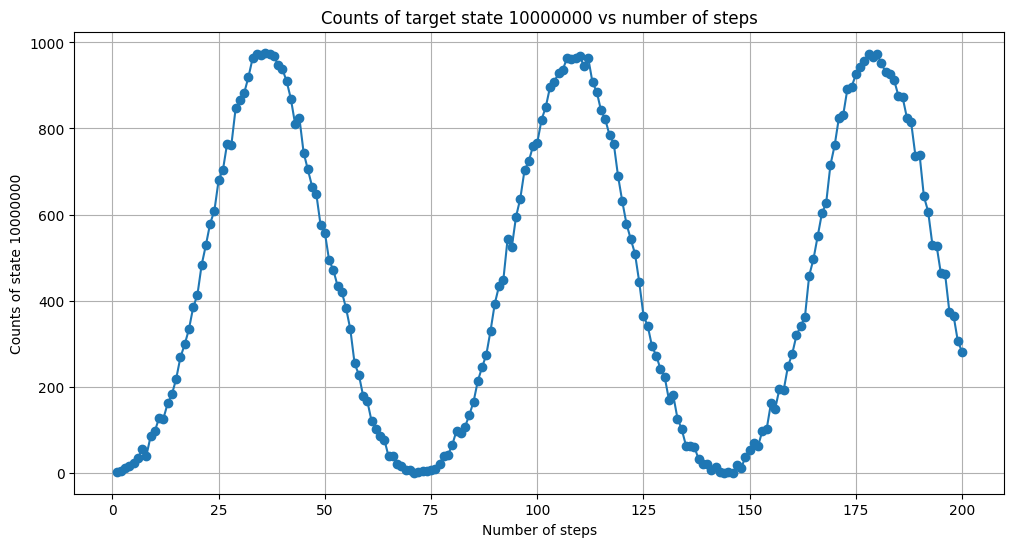"}
        \caption{Marked state probability over walk steps.}
        \label{fig:16x16_prob}
    \end{subfigure}
    \caption{Ideal noiseless results for a $16\times 16$ grid with a single marked vertex.}
    \label{fig:grid_16x16}
\end{figure}

\begin{figure}[htpb]
    \centering
    \begin{subfigure}[b]{0.48\linewidth}
        \centering
        \includegraphics[width=\linewidth]{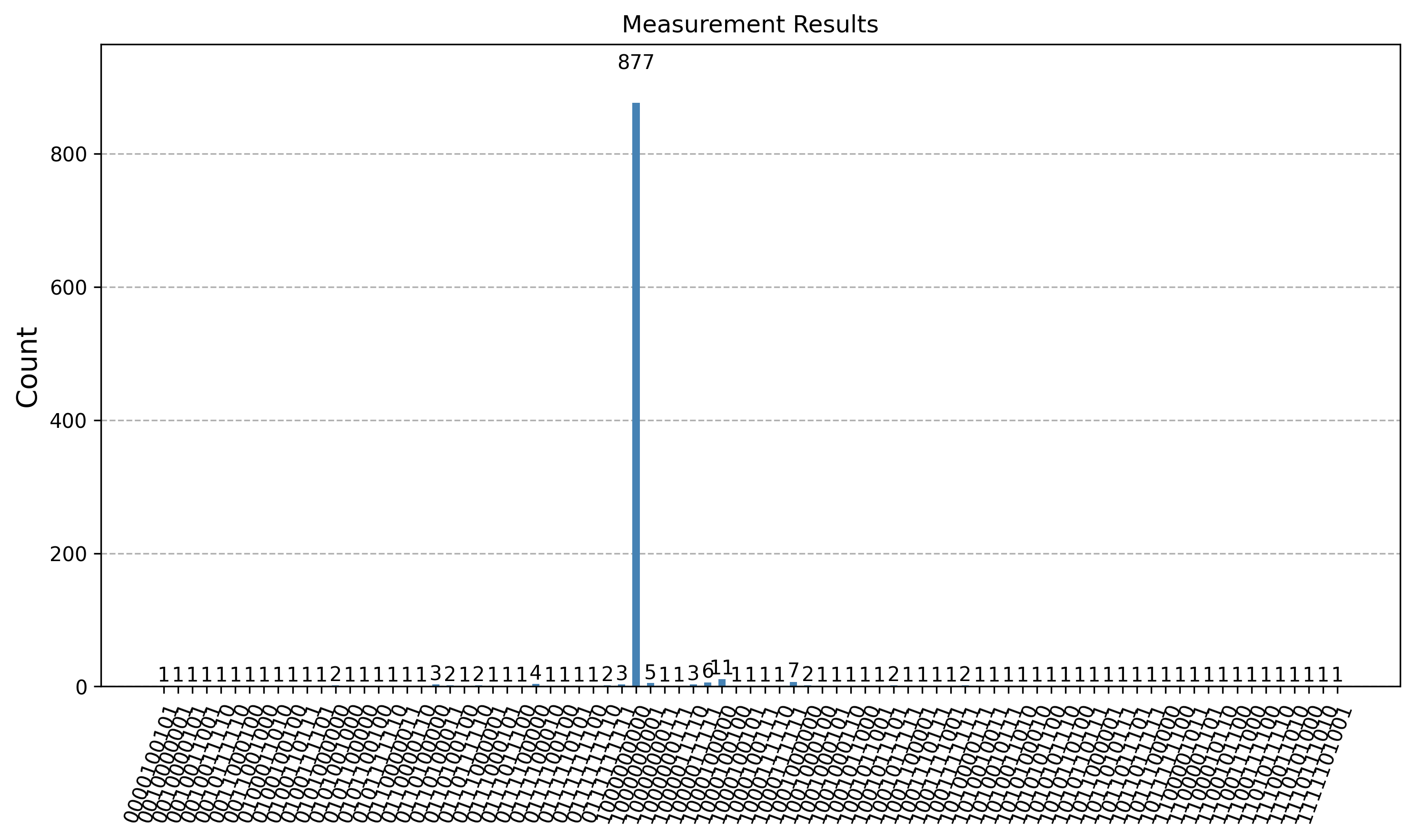}
        \caption{Final measurement distribution after $\sqrt{N\log_2 N}$ walk steps.}
        \label{fig:32x32_hist}
    \end{subfigure}
    \hfill
    \begin{subfigure}[b]{0.48\linewidth}
        \centering
        \includegraphics[width=\linewidth]{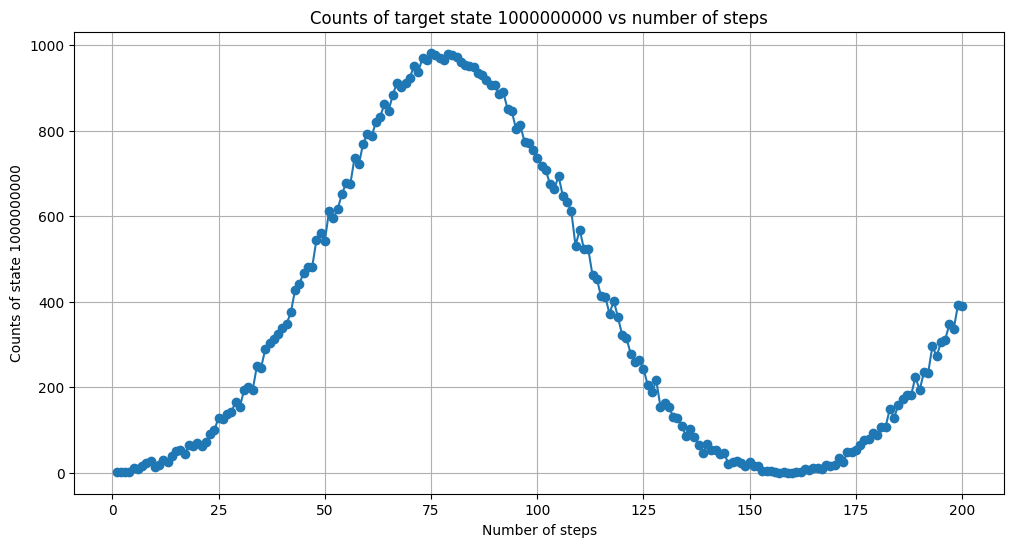}
        \caption{Marked state probability over walk steps.}
        \label{fig:32x32_prob}
    \end{subfigure}
    \caption{Ideal noiseless results for a $32\times 32$ grid with a single marked vertex.}
    \label{fig:grid_32x32}
\end{figure}

\begin{figure}[htpb]
    \centering
    \begin{subfigure}[b]{0.48\linewidth}
        \centering
        \includegraphics[width=\linewidth]{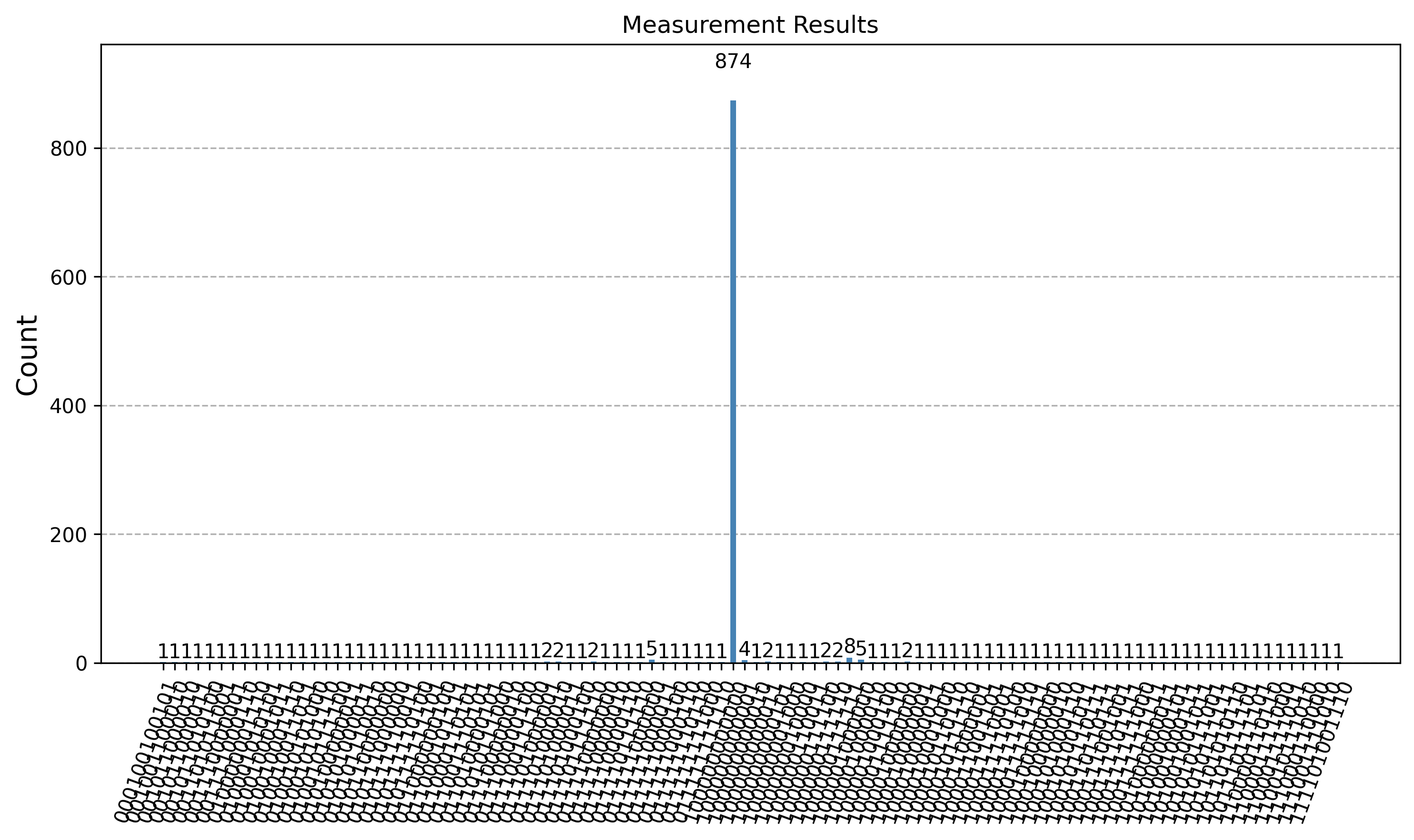}
        \caption{Final measurement distribution after $\sqrt{N\log_2 N}$ walk steps.}
        \label{fig:64x64_hist}
    \end{subfigure}
    \hfill
    \begin{subfigure}[b]{0.48\linewidth}
        \centering
        \includegraphics[width=\linewidth]{"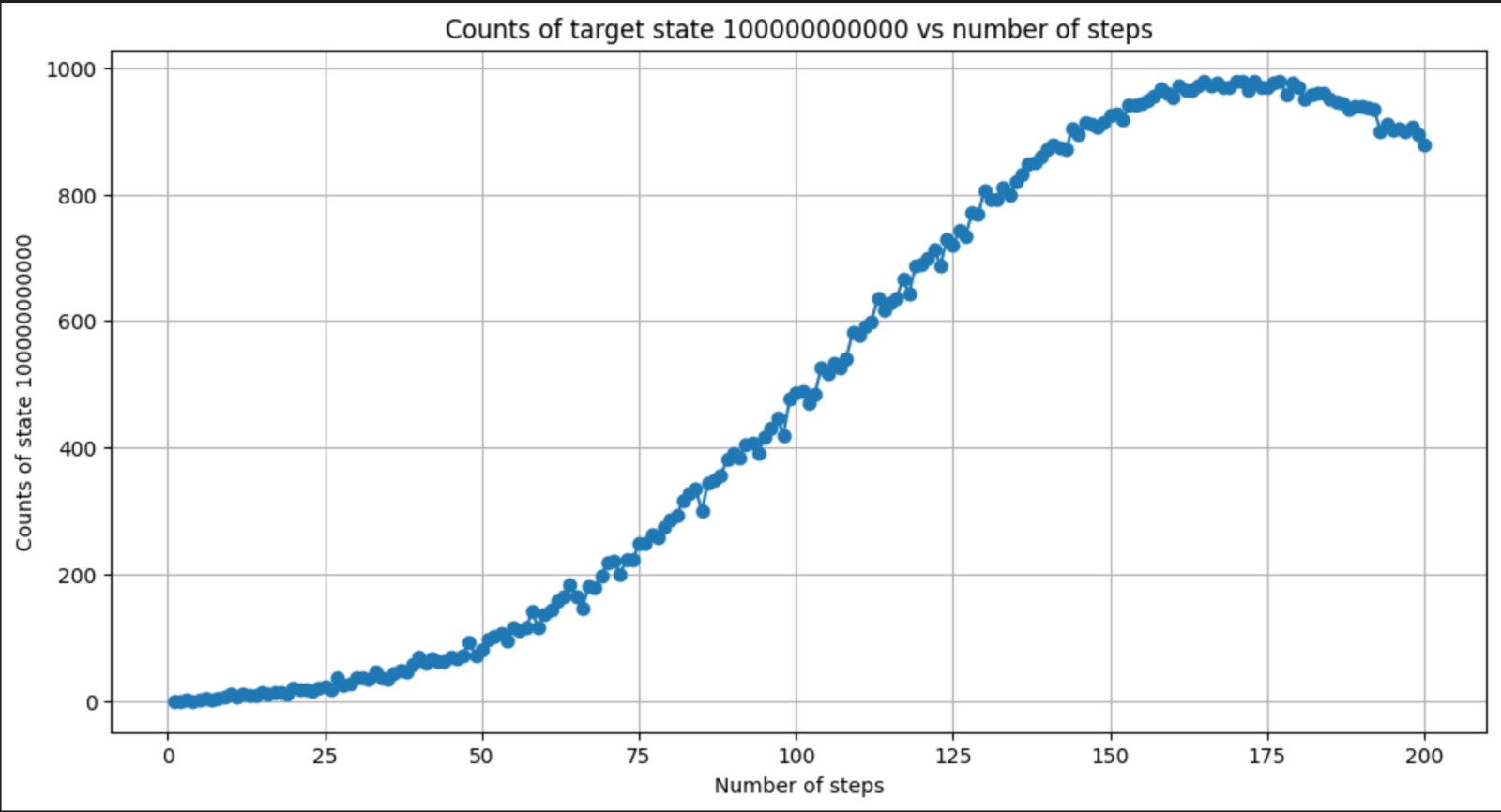"}
        \caption{Marked state probability over walk steps.}
        \label{fig:64x64_prob}
    \end{subfigure}
    \caption{Ideal noiseless results for a $64\times 64$ grid with a single marked vertex.}
    \label{fig:grid_64x64}
\end{figure}

Figures~\ref{fig:grid_8x8}--\ref{fig:grid_64x64} show the results for increasing grid sizes. In each figure, the histogram displays the final measurement distribution after the selected number of walk steps, while the probability curve tracks the marked-state probability across the walk evolution. The marked vertex receives the dominant probability mass at the expected peak, while the unmarked states remain comparatively suppressed. The step-wise probability curves also show the characteristic oscillatory behavior of quantum walk search, where the success probability increases up to an optimal measurement time and then decreases as the walk continues. These observations confirm that the simulated circuit reproduces the expected amplitude amplification behavior of lackadaisical quantum walk search \cite{wong2018faster}.

\begin{comment}

We have implemented the lackadaisical quantum walks on $\sqrt{N} \times \sqrt{N}$ grids 
ranging for $8 \times 8$, with a single marked vertex. The 
walks is initialized in uniform superposition with self-loop weight $\ell = 4/N$, 
and measured over $1000$ trials after $\sqrt{N \log_2 N}$ steps. Across all grid sizes, the marked state is identified with consistently high success probability, while all other states receive negligible counts. The step-wise probability plots 
exhibit characteristic oscillatory behavior, with the oscillation period scaling 
as $\mathcal{O}(\sqrt{N})$, confirming the $\mathcal{O}(\sqrt{N\log N})$ query Complexity of the lackadaisical quantum walks on 2D grids. The following illustrations demonstrate the algorithm in [Fig 1,2,3,4]
\end{comment}

\newpage
\subsubsection{Multiple marked states} 

We next evaluate the circuit for the case of multiple marked vertices. For this experiment, we use an $8\times 8$ grid and mark two vertices, $|100000\rangle$ and $|111100\rangle$. The oracle applies a phase flip to both marked vertices, and the success probability is computed as the total probability of measuring either target.

\begin{figure}[!h]
    \centering
    \begin{subfigure}[]{0.48\linewidth}
        \centering
        \includegraphics[width=\linewidth]{"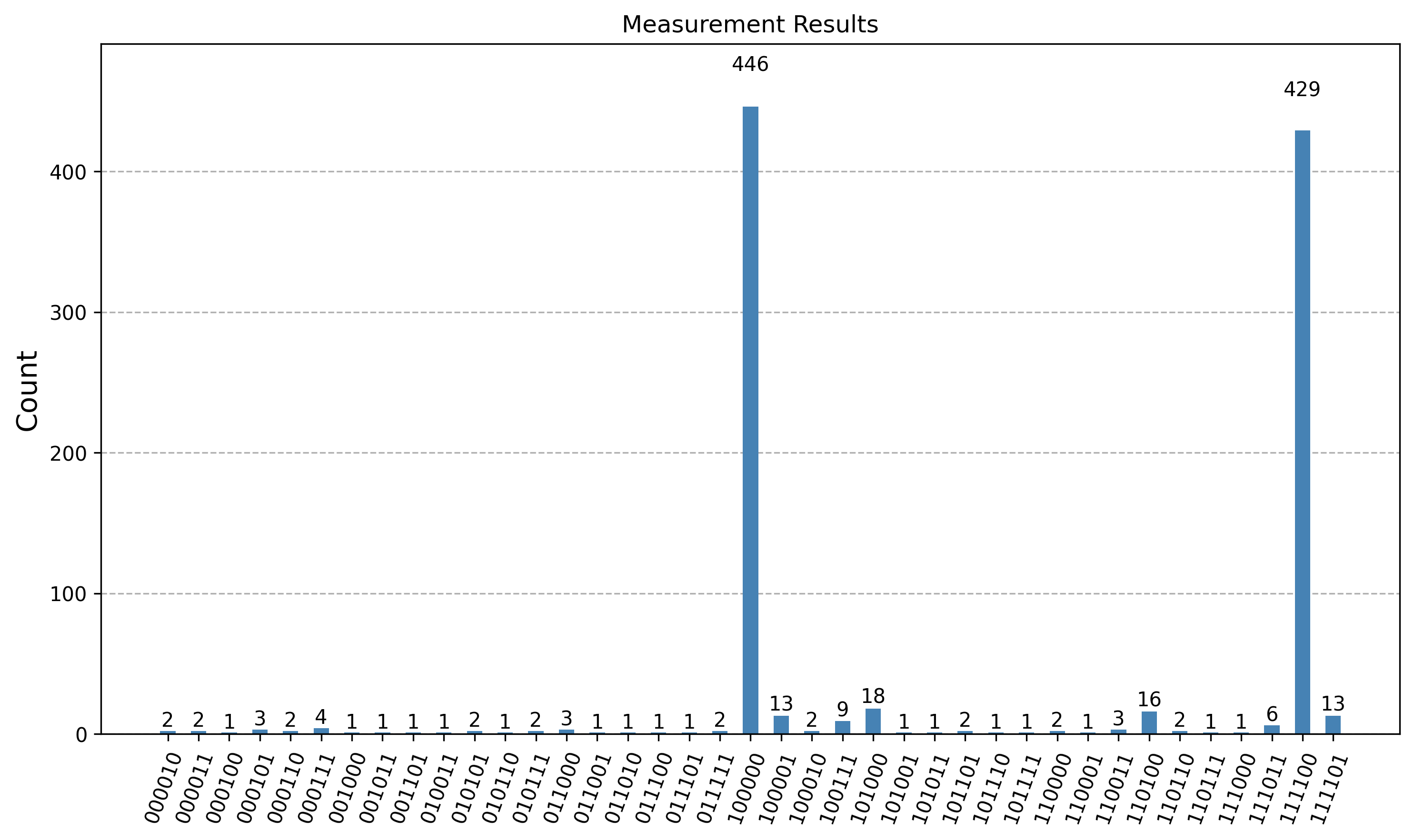"}
        \caption{Final measurement distribution after $\sqrt{\frac{N}{2}\log_2 \frac{N}{2}}$ steps.}
        \label{fig:multi_hist}
    \end{subfigure}
   \hfill
    \begin{subfigure}[]{0.48\linewidth}
        \centering
        \includegraphics[width=\linewidth]{"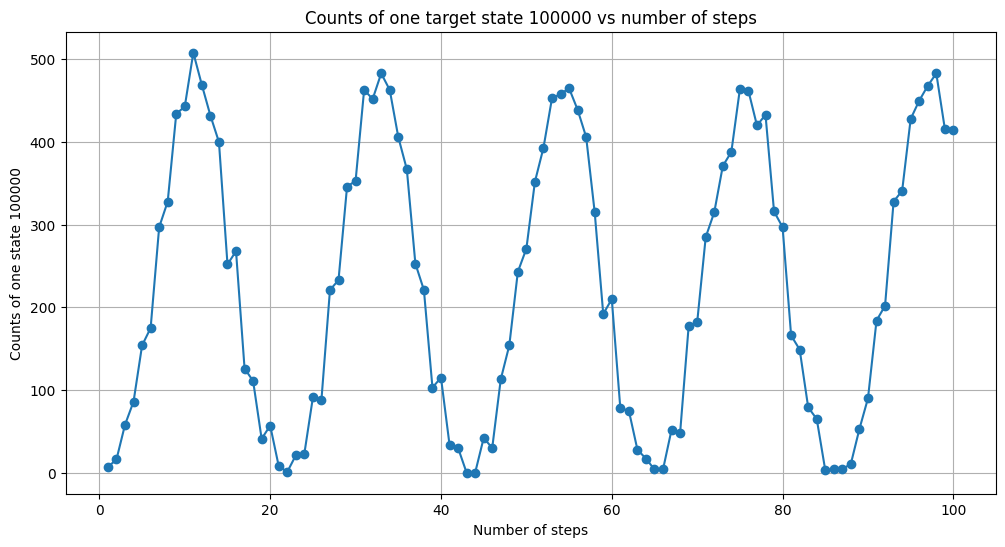"}
        \caption{Marked state probability over walk steps.}
        \label{fig:multi_prob}
    \end{subfigure}
    \caption{Ideal noiseless results for an $8\times 8$ grid with two marked vertices, $|100000\rangle$ and $|111100\rangle$.}
    \label{fig:grid_8x8_multi}
\end{figure}

Figure~\ref{fig:grid_8x8_multi} shows the resulting distribution and the evolution of the probability. The step-wise probability curve retains the oscillatory behavior observed in the single marked vertex experiment, indicating that the circuit naturally extends to the multi-target setting. These results are consistent with the expected behavior of lackadaisical quantum walk search with multiple marked vertices~\cite{saha2018searchclusteredmarkedstates, doi:10.1142/S0129054118410113, Saha_2022, giri2020lackadaisical}.

%We extend the evaluation to the case of multiple marked vertices on a $\sqrt{N} \times \sqrt{N}$ grid With two marked vertices, the success probability is distributed across both targets. The measurement distributions (subfigure (a)) show two dominant peaks, one at each marked state, with their combined counts reflecting a high total success probability. The step-wise plots (subfigure (b)) retain the oscillatory behavior seen in the single marked case, though the per-target probability is reduced by approximately half. This confirms that the lackadaisical quantum walks scales naturally to multiple targets while preserving its $\mathcal{O}\left(\sqrt{\frac{N}{M}\log \frac{N}{M}}\right)$ query complexity\cite{giri2020lackadaisical}.

\subsection{Noisy Simulation}
\label{sec:noisy}

The ideal simulations verify the correctness of the circuit construction, but they do not capture the behavior of the implementation under realistic hardware constraints. We therefore evaluate the proposed circuit using a noise model inspired by the superconducting device. In particular, we use the calibrated noise model of the IBM Fez backend available through Qiskit. The model includes relaxation and dephasing effects, gate errors, and readout errors.

The main noise sources considered are summarized as follows:
\begin{itemize}
\item \textbf{Decoherence:} Energy relaxation and dephasing, characterized by $T_1$ and $T_2$, cause quantum states to decay during circuit execution.
\item \textbf{Gate errors:} Imperfect single-qubit and two-qubit operations introduce deviations from the intended unitary gates.
\item \textbf{Readout errors:} Measurement errors cause the observed classical bit string to differ from the actual final quantum state.
\end{itemize}

Table~\ref{tab:ibm_fez_noise} lists the main backend parameters used in the noisy simulations.

%The robustness of the lackadaisical quantum walks under realistic settings was investigated by extending our experiments to a noisy setting by using the noise model of a realistic IBM quantum processing unit (QPU). In particular, we made use of the calibrated noise model of \textbf{IBM Fez}, which is a superconducting 156-qubit quantum computer designed using Heron~r2 architecture, available via the Qiskit Runtime framework. The noise model includes:

\begin{comment}

\begin{itemize}
    \item \textbf{Decoherence}: energy relaxation ($T_1$) and 
    dephasing ($T_2$) causes the qubit state to decay toward the 
    ground state over time.
    
    \item \textbf{Gate errors}: imperfect unitary operations introduce depolarizing noise on both single-qubit and two-qubit gates.
    
    \item \textbf{Readout errors}: measurement operations are 
    subject to bit-flip errors, causing the observed outcome to 
    differ from the true qubit state.
\end{itemize}

Table~\ref{tab:ibm_fez_noise} summaries the key noise parameters 
of the IBM Fez device used in our simulation, as obtained from the 
Qiskit backend calibration data.
\end{comment}
\begin{table}[H]
    \centering
    \caption{Noise profile of the IBM Fez backend used for noisy simulation.}\renewcommand{\arraystretch}{1.3}
    \begin{tabular}{lcc}
        \toprule
        \textbf{Parameter}                  & \textbf{Symbol}   & \textbf{Value} \\
        \midrule
        Number of qubits                    & $n_q$             & 156 \\
        Processor architecture              & ---               & Heron r2 \\
        Median $T_1$ (relaxation time)      & $T_1$             & 141.07 $\mu$s \\
        Median $T_2$ (dephasing time)       & $T_2$             & 102.05 $\mu$s \\
        Median readout error                & $\epsilon_{ro}$   & 1.190\% \\
        \bottomrule
    \end{tabular}
    
    \label{tab:ibm_fez_noise}
\end{table}

\begin{figure}[!h]
    \centering
    \includegraphics[width=0.5\linewidth]{"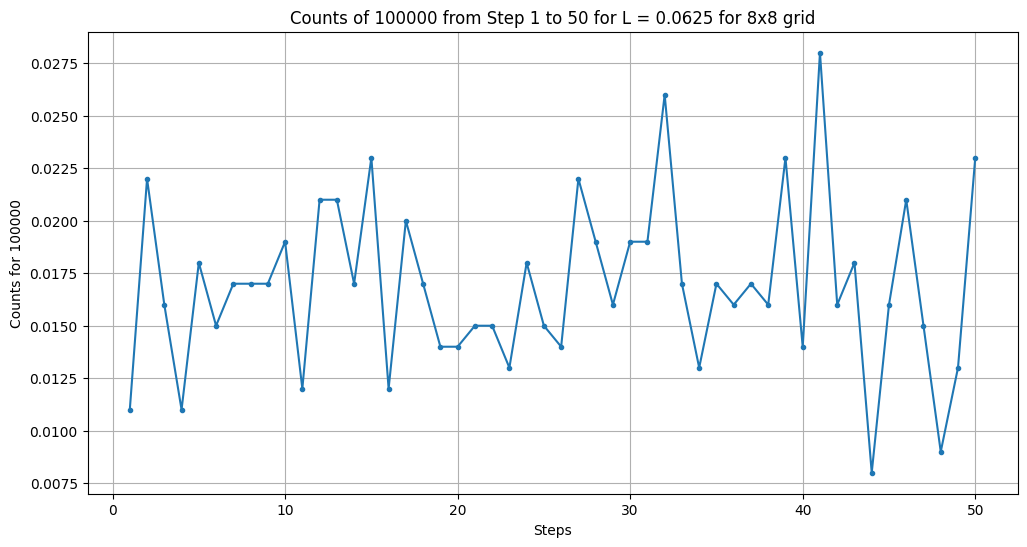"}
    \caption{Noisy simulation result for an $8\times 8$ grid using the IBM Fez noise model.}
    \label{fig:noise_8x8}
\end{figure}

Figure~\ref{fig:noise_8x8} shows the noisy result for the $8\times 8$ grid. Compared with the ideal noiseless case, the marked-state probability is significantly reduced and the clean oscillatory pattern becomes damped. This behavior is expected, since quantum walk search relies on coherent interference to amplify the marked vertex. Noise disrupts this interference and reduces the effectiveness of the search process.

These noisy outputs show the influence of quantum noise on the amplification of the probability of the marked state. Consequently, in order to enhance the reliability of the achieved outputs and regain some quantum information which was lost because of noise, we apply some methods for noise reduction. In addition, we change the value of the noise parameter $\ell$ to see the efficiency of the employed noise reduction technique under various noise conditions.

To improve the observed performance in this noisy setting, we consider two complementary strategies: first, applying noise-mitigation techniques to reduce the bias in the measured output distribution, and second, varying the self-loop weight $\ell$ to examine whether parameter tuning can recover a higher marked-state success probability under noise. The results of these two approaches are discussed in the following subsections.

\subsubsection{Noise mitigation}
To improve the reliability of the noisy results, we first apply several noise-mitigation and error-suppression techniques. These methods do not implement full quantum error correction; instead, they aim to reduce the bias introduced by noise or extrapolate the measured result toward the noiseless limit.

The mitigation techniques considered in this work are:  (a) Zero-noise extrapolation (ZNE): The circuit noise is artificially amplified, for example through gate folding, and the measured values are extrapolated back to the zero-noise limit~\cite{koenig2024},
 (b) Dynamical decoupling (DD): Additional pulse sequences are inserted during idle periods to suppress decoherence effects~\cite{khan2024},
 (c) Twirled readout error extinction (TREX): Randomized bit flips and calibration data are used to reduce readout bias~\cite{pomarico2025},
 (d) Probabilistic error amplification (PEA): The noise level is amplified in a controlled manner using a learned noise model, enabling more accurate zero-noise extrapolation~\cite{koenig2024},
(e) Probabilistic error cancellation (PEC): The ideal operation is represented as a quasi-probability combination of noisy implementable operations, reducing noise bias at the cost of increased sampling overhead~\cite{song2019}.

\begin{figure}[!h]
    \centering
    \begin{subfigure}[b]{0.48\linewidth}
        \centering
        \includegraphics[width=\linewidth]{"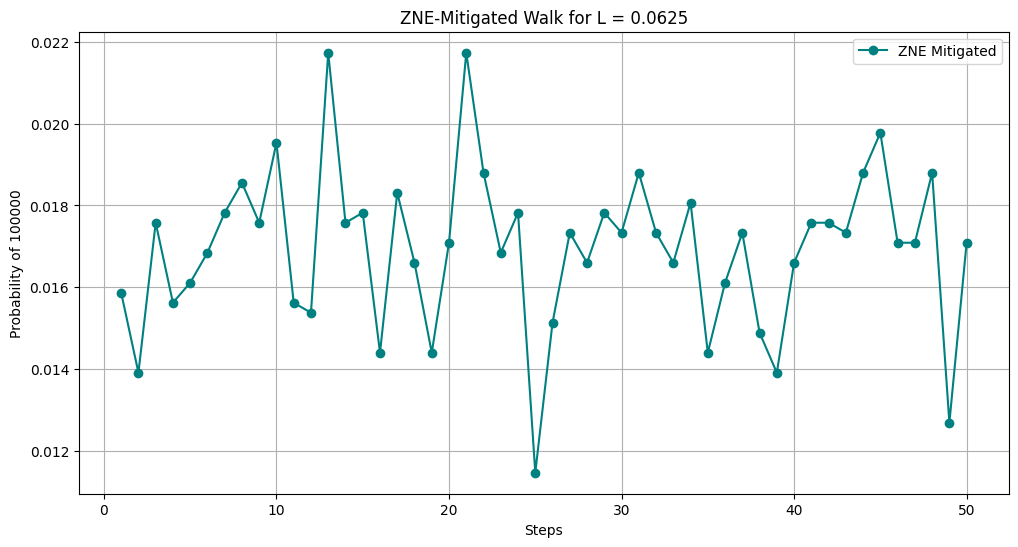"}
        \caption{Zero-Noise Extrapolation (ZNE)}
        \label{fig:noise_zne}
    \end{subfigure}
    \hfill
    \begin{subfigure}[b]{0.48\linewidth}
        \centering
        \includegraphics[width=\linewidth]{"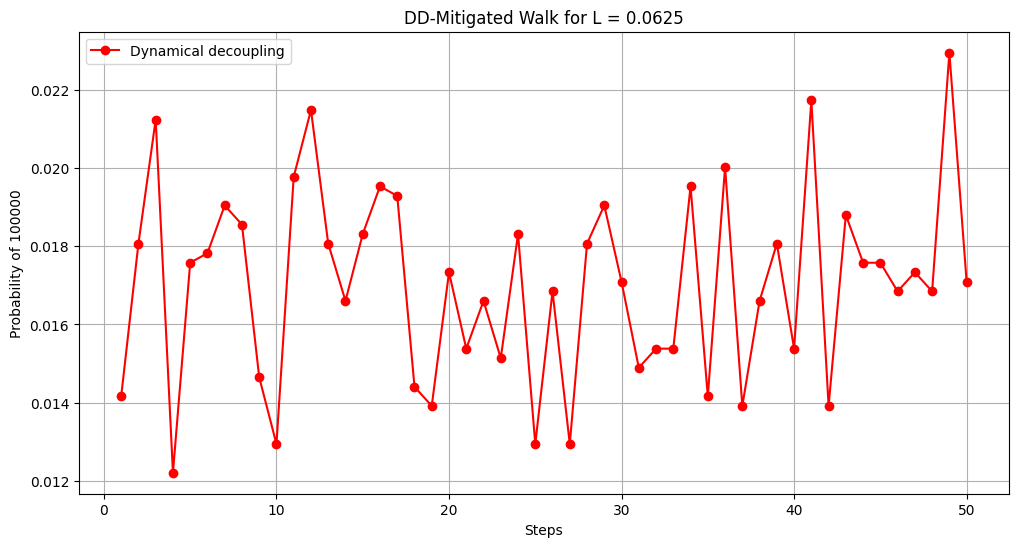"}
        \caption{Dynamical Decoupling (DD)}
        \label{fig:noise_dd}
    \end{subfigure}
    
    \vspace{1em} 
    
    \begin{subfigure}[b]{0.48\linewidth}
        \centering
        \includegraphics[width=\linewidth]{"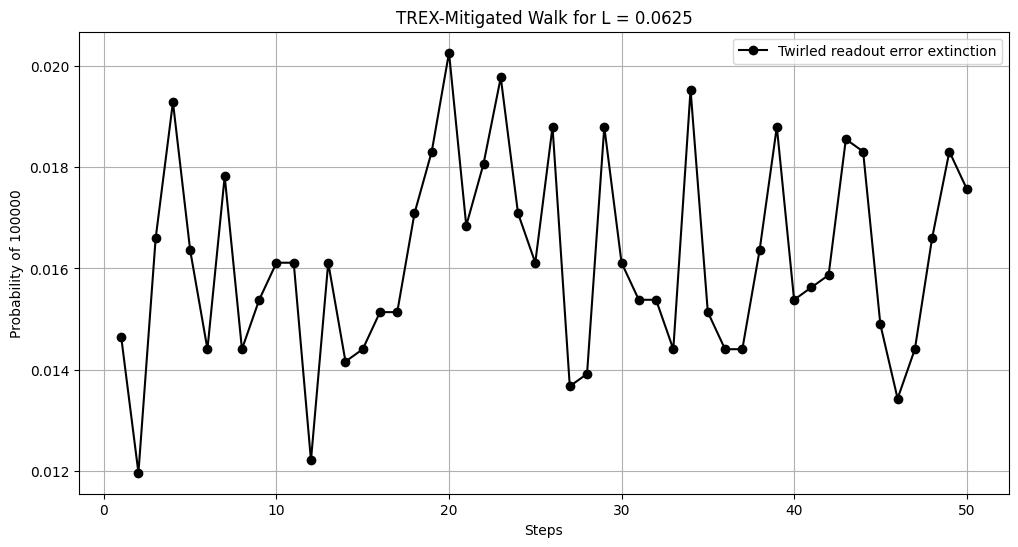"}
        \caption{Twirled Readout Error eXtinction (TREX)}
        \label{fig:noise_trex}
    \end{subfigure}
    \hfill
    \begin{subfigure}[b]{0.48\linewidth}
        \centering
        \includegraphics[width=\linewidth]{"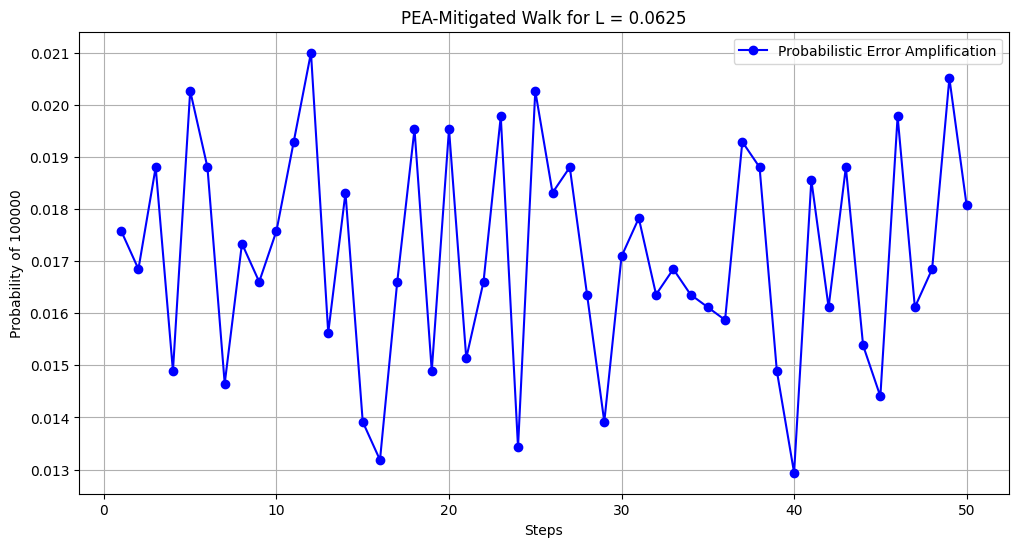"}
        \caption{Probabilistic Error Amplification (PEA)}
        \label{fig:noise_pea}
    \end{subfigure}
    
    \vspace{1em} 
    
    \begin{subfigure}[b]{0.48\linewidth}
        \centering
        \includegraphics[width=\linewidth]{"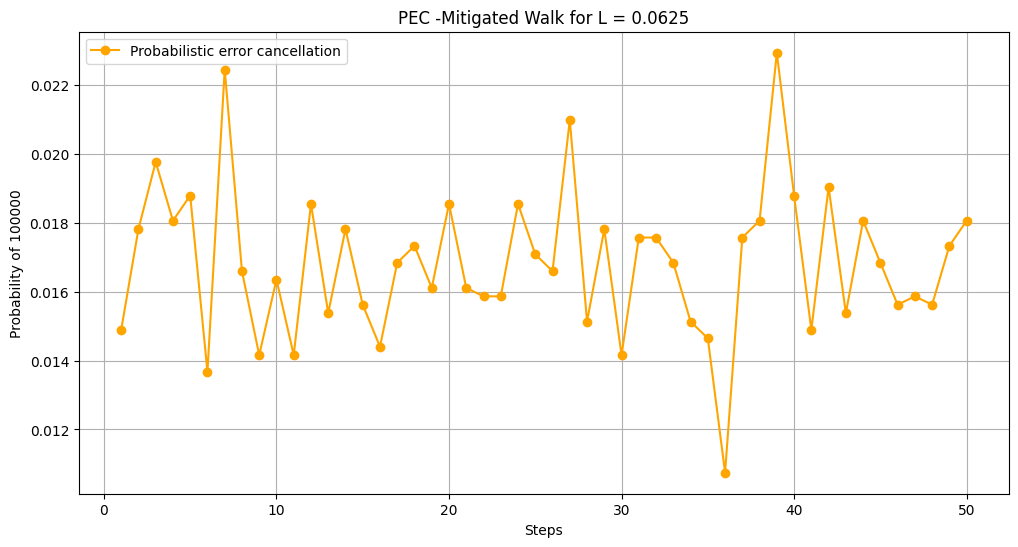"}
        \caption{Probabilistic Error Cancellation (PEC)}
        \label{fig:noise_pec}
    \end{subfigure}
    
    \caption{Marked state probability for an $8\times 8$ grid under various noise mitigation techniques.}
    \label{fig:noise_mitigation_comparison}
\end{figure}

Figure~\ref{fig:noise_mitigation_comparison} compares the marked-state probability obtained after applying the considered mitigation techniques. The results show that these methods do not recover the marked state amplification observed in the ideal simulation. In fact some cases, the mitigated distributions remain close to the unmitigated noisy result, indicating that the accumulated gate errors and circuit depth dominate the error profile. Therefore, while noise mitigation provides a useful diagnostic tool for assessing the noisy circuit behavior, it is not sufficient by itself to restore the ideal lackadaisical quantum walk dynamics in the present implementation.

\subsubsection{Effect of varying the self-loop weight $\ell$ under noise}

Since the mitigation techniques do not recover the ideal marked state amplification, we further investigate whether the self-loop weight $\ell$ can be tuned to improve the noisy search behavior. The parameter $\ell$ controls the balance between propagation across the grid and localization at the current vertex. Since the noisy experiments are performed on an $8\times 8$ grid, the total number of vertices is $N=64$, and the theoretically motivated self-loop weight is $\ell=4/N=0.0625$ in the ideal setting. To examine whether this value remains effective under noise, we choose a set of self-loop weights below, near, and above $0.0625$:
\begin{equation}
\ell \in {0.010,\ 0.025,\ 0.050,\ 0.075,\ 0.100}.
\end{equation}
This range allows us to test whether weaker or stronger self-loop localization can improve the marked state probability when the circuit is affected by gate noise, decoherence, and readout errors.

\begin{figure}[!h] 
    \centering
    
    \begin{subfigure}[b]{0.48\linewidth}
        \centering
        \includegraphics[width=\linewidth]{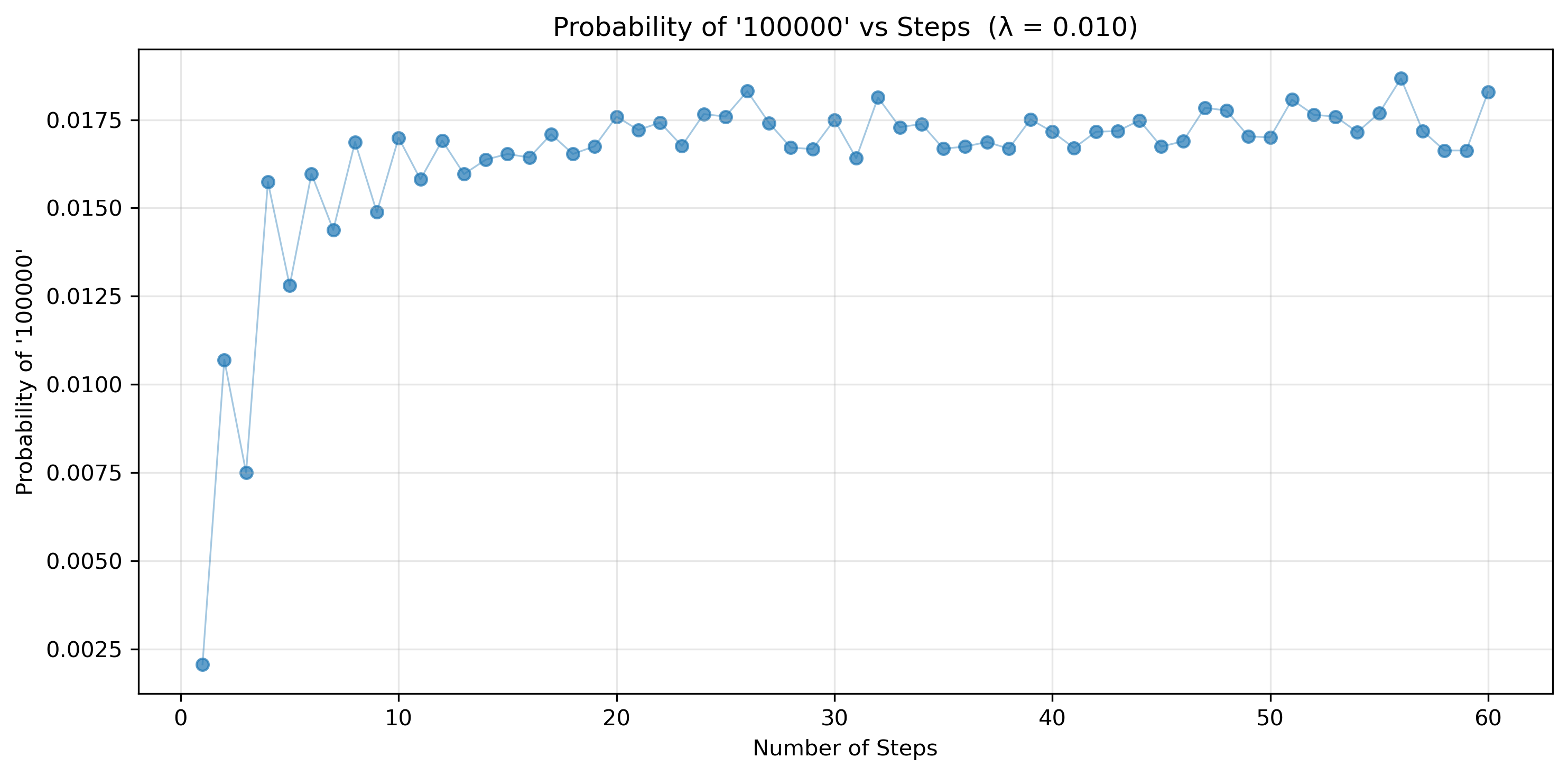} 
        \caption{$\ell = 0.010$}
        \label{fig:scatter_010}
    \end{subfigure}
    \hfill
    \begin{subfigure}[b]{0.48\linewidth}
        \centering
        \includegraphics[width=\linewidth]{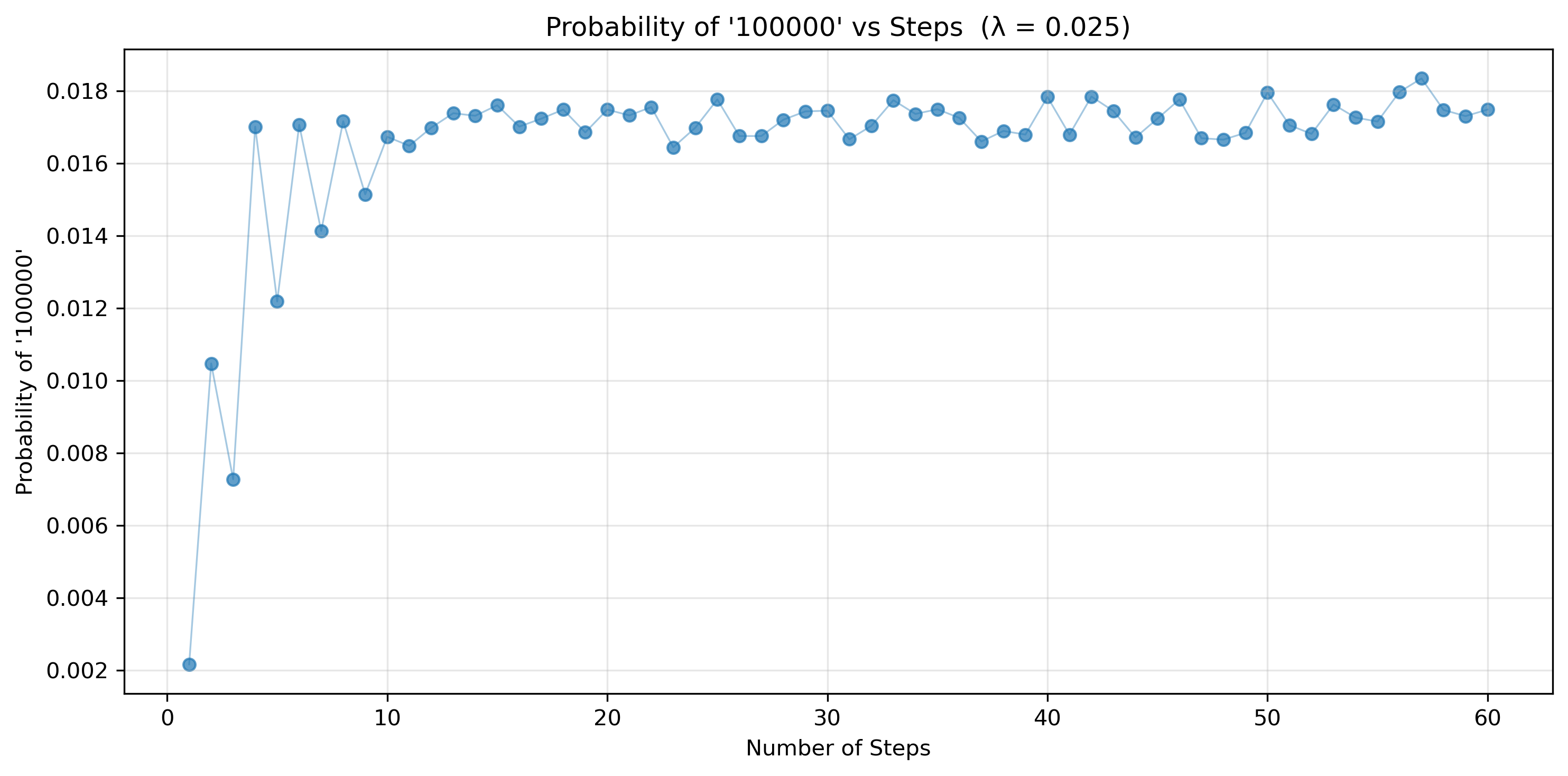}
        \caption{$\ell = 0.025$}
        \label{fig:scatter_025}
    \end{subfigure}
    
    \vspace{1em} 
    
    \begin{subfigure}[b]{0.48\linewidth}
        \centering
        \includegraphics[width=\linewidth]{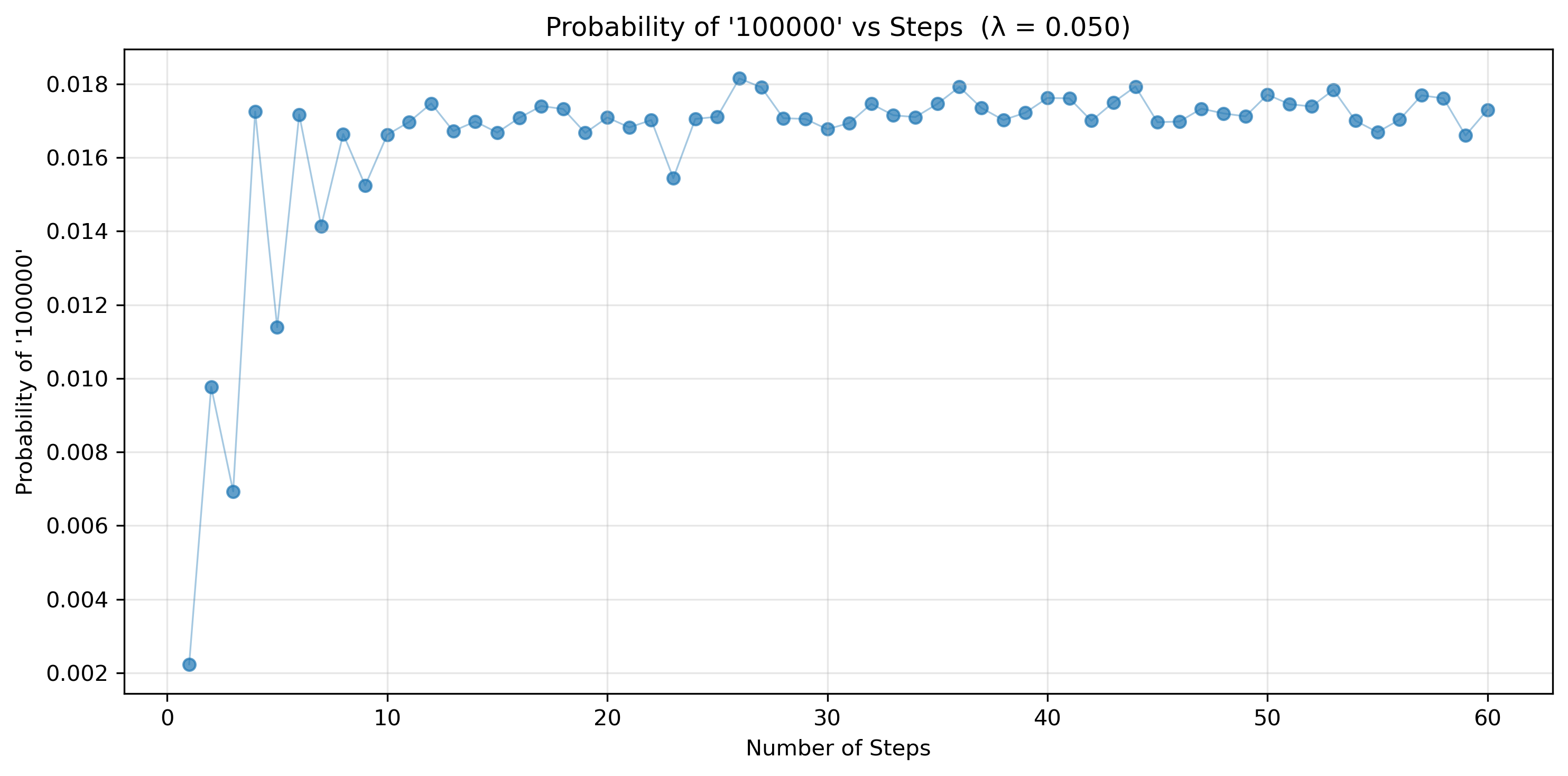}
        \caption{$\ell = 0.050$}
        \label{fig:scatter_050}
    \end{subfigure}
    \hfill
    \begin{subfigure}[b]{0.48\linewidth}
        \centering
        \includegraphics[width=\linewidth]{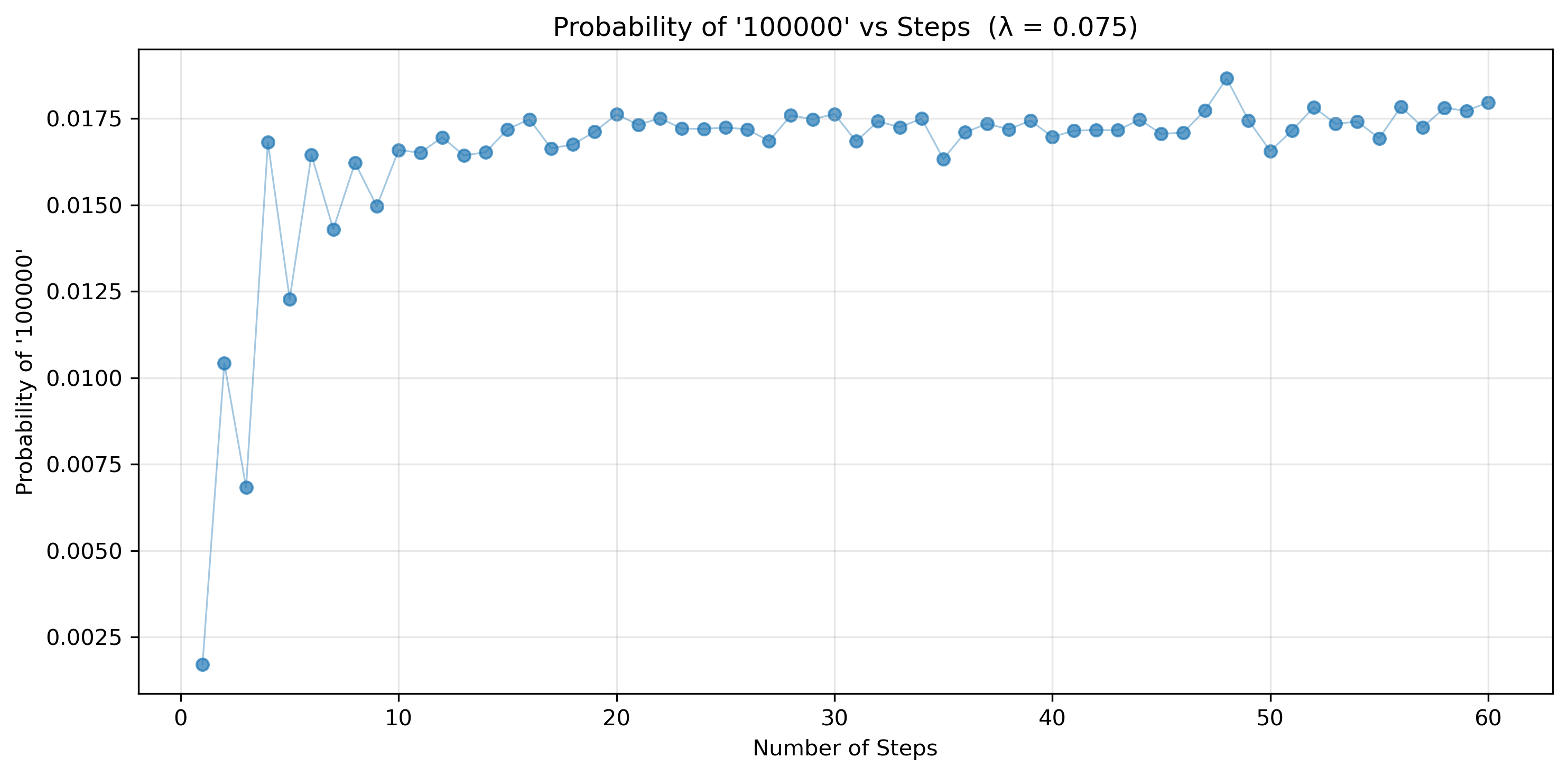}
        \caption{$\ell = 0.075$}
        \label{fig:scatter_075}
    \end{subfigure}
    
    \vspace{1em} 
    
    \begin{subfigure}[b]{0.48\linewidth}
        \centering
        \includegraphics[width=\linewidth]{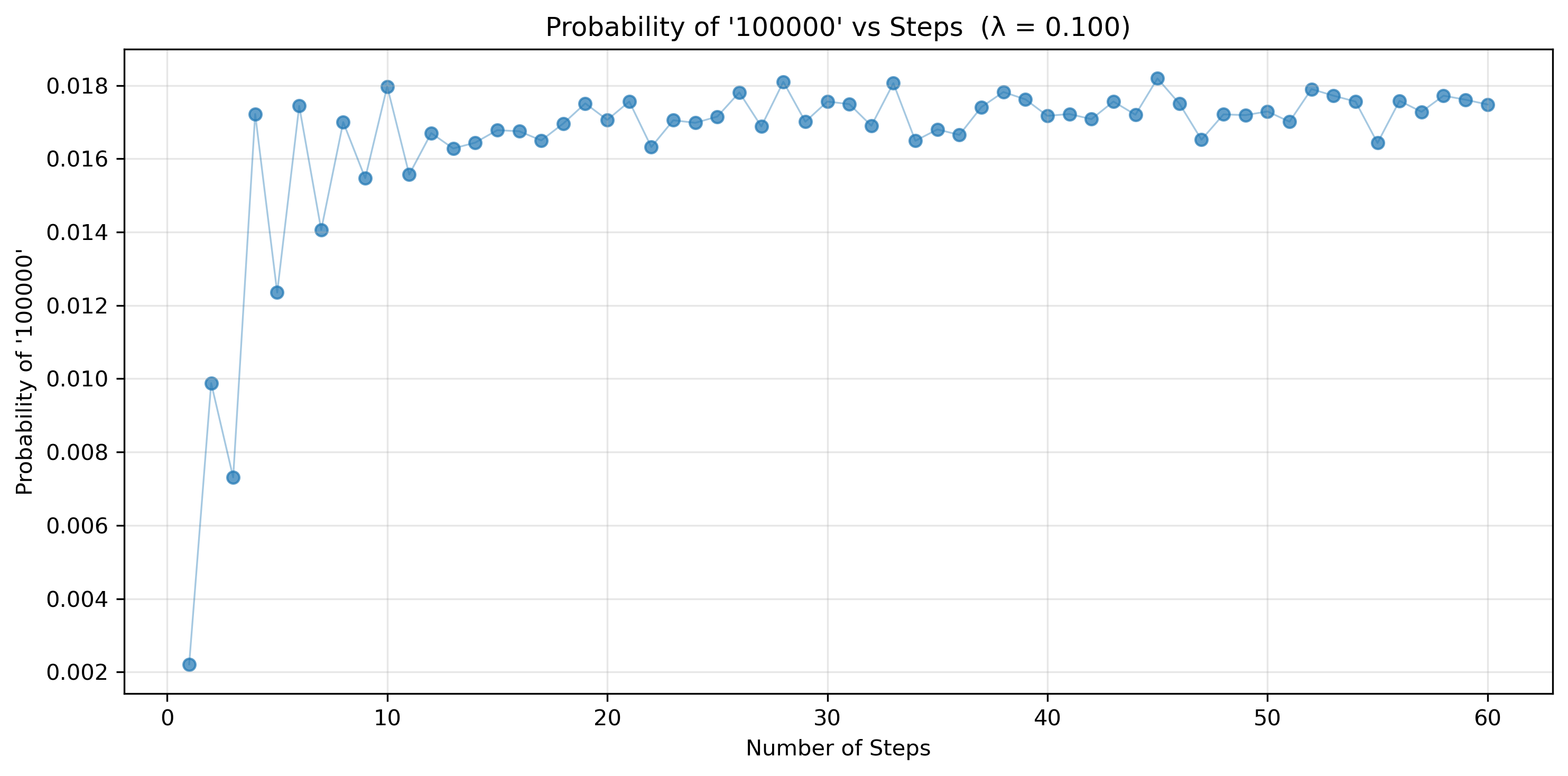}
        \caption{$\ell = 0.100$}
        \label{fig:scatter_100}
    \end{subfigure}
    
    \caption{Marked state probability in an $8\times 8$ grid under noisy simulation for different self-loop weights $\ell$.}
    \label{fig:scatter_noise_L}
\end{figure}

Figure~\ref{fig:scatter_noise_L} shows the marked state probability for the tested values of $\ell$ on an $8\times 8$ grid under noisy simulation. Across these simulations, the marked state probability initially increases during the early walk steps, but the coherent oscillatory behavior observed in the ideal case is damped by noise. The results indicate that varying $\ell$ changes the noisy search profile, but the improvement remains limited for the tested range. This suggests that self-loop tuning can serve as an additional empirical parameter for noisy implementations, although it cannot fully compensate for the accumulated errors caused by circuit depth and imperfect gates.

\begin{comment}

Varying the noise parameter ($\ell$) is essential for assessing the robustness of the lackadaisical quantum walk search algorithm under realistic quantum computing conditions. In practical NISQ devices, noise levels are not fixed and may vary depending on the hardware platform, circuit depth, qubit connectivity, and environmental interactions. By analyzing a range of noise strengths
$\ell \in \{0.010,\ 0.025,\ 0.050,\ 0.075,\ 0.100\}$. As shown in 
Figures~\ref{fig:scatter_010}--\ref{fig:scatter_100}, the probability of the marked 
state rises sharply in the first few steps, undergoes brief oscillations, and then 
stabilizes to a steady-state value of approximately $0.016$--$0.018$ across all 
noise levels. Notably, the final steady-state probability remains consistent 
regardless of $\ell$, demonstrating that the lackadaisical quantum walks is remarkably 
robust to varying noise intensities. This suggests that moderate depolarizing noise 
does not significantly degrade the algorithm's ability to locate the marked vertex.
\end{comment}

Overall, the numerical results verify that the proposed circuit reproduces the expected lackadaisical quantum walk search behavior in the ideal setting and naturally extends to multiple marked vertices. Under realistic noise, however, the marked-state amplification is substantially degraded. The noise-mitigation experiments show limited recovery, and the self-loop tuning experiments indicate that varying $\ell$ can modify the noisy search profile but does not fully restore the ideal dynamics. These observations highlight the importance of circuit optimization, resource analysis, and fault-tolerant considerations, which are discussed in the next section.

%Even after implementing noise mitigation techniques and thoroughly analyzing various levels of the noise parameter $\ell$, the recovery of search efficacy remains constrained by the fundamental physical limitations of NISQ devices. These results clarify that the current performance bottleneck originates from contemporary hardware constraints rather than the proposed circuit structure. Consequently, our gate-level synthesis should be contextualized as a future-oriented framework intended for fault-tolerant quantum computers. To properly evaluate its viability in this future paradigm, a comprehensive resource analysis is essential.
\section{Resource Analysis}
\label{sec:resource}

The proposed lackadaisical quantum walk search algorithm is designed as a gate-level circuit. Therefore, beyond verifying the search behavior, it is important to quantify the resources required for its implementation. This section analyzes the resource cost at two levels. First, we report the logical circuit resources, including gate counts, circuit depth, and non-Clifford depth for different grid sizes. Second, we estimate the physical resources required under a surface code-based fault-tolerant implementation model. This will provide us an understanding how realistic it will be to run on future large-scale fault-tolerant quantum computers.

For an $L\times L$ grid with $L=2^n$, the position of the walker is encoded using two $n$-qubit registers, and the lackadaisical coin is encoded using three additional qubits. Thus, the number of algorithmic qubits is
\begin{equation}
n_{\mathrm{alg}} = 2n+3 = \log_2 N + 3,
\end{equation}
where $N=L^2$ is the total number of grid vertices. This count includes only the logical qubits required by the algorithmic registers and does not include additional ancilla qubits that may be introduced by a particular gate decomposition or fault-tolerant compilation procedure.

\begin{comment}

Since the primary contribution of this work is the development of a gate-level design of the lackadaisical quantum walk search algorithm, it is necessary to analyze the amount of resources consumed by the implementation method proposed.The resource analysis serves as an assessment of the difficulty of implementing this approach, as well as a tool for understanding how realistic it will be to run on future large-scale fault-tolerant quantum computers. This section is divided into two sub-sections: logical analysis and physical analysis. The implementation section focuses on determining the best trade-off among speed, error correction, and cost between quantum algorithm theory and the physical system.
\end{comment}
\subsection{Logical resource analysis}

All circuit resource metrics were obtained using Qiskit 2.3.1 with Qiskit Aer 0.17.2. Circuits were transpiled with a fixed transpiler seed of 42 to ensure reproducibility. Multi-controlled gate operations were synthesized using the noancilla decomposition, without introducing clean or dirty ancillary qubits. No device-specific coupling map was imposed; consequently, the reported circuit depth represents the logical dependency depth of the transpiled circuit, rather than a hardware-scheduled execution depth obtained from ASAP/ALAP scheduling. Gate counts were evaluated after transpilation into the simulator-supported basis, while the global phase tracked by Qiskit was not included as a physical resource. These settings provide a consistent basis for comparing the logical resource requirements across different grid sizes.

Table~\ref{tab:logical} reports the logical gate counts for the 
lackadaisical quantum walk search circuit for grid sizes ranging from $8\times 8$ to $64\times 64$. The decomposition includes two-qubit gates ($CX$), single-qubit gates ($X$, $Z$, $S$, $S^\dagger$, $T$, $T^\dagger$, $H$), along with measurement operations. The results show that the complete circuit depth grows from $213,156$ for the $8\times 8$ grid to $2,555,755$ for the $64\times 64$ grid. The non-Clifford depth also increases substantially, from $73,352$ to $1,007,102$.

%The total logical gate count grows approximately quadratically with the linear grid dimension. This growth is expected because larger grids require more position qubits and more complex controlled increment, decrement, oracle, and shift operations. In particular, the $H$ count, the $CX$ count and the non-Clifford gate count increase significantly as the grid size grows. Since non-Clifford gates, especially $T$ and $T^\dagger$ gates, are expensive in fault-tolerant quantum computing due to magic-state distillation, the non-Clifford circuit depth is a key indicator of the cost of scalable implementation.

%In fault-tolerant quantum computing, Clifford gates (H, CX, S) can be implemented transversally and are relatively inexpensive. Non-Clifford gates, particularly the T and T$^\dagger$ gates, cannot be implemented  transversally and require costly magic state distillation protocols. Consequently, the non-Clifford gate depth serves as the primary bottleneck determining the true resource cost of a fault-tolerant implementation. Tracking this depth separately from the total circuit depth provides a more accurate estimate of the practical overhead required for error-corrected execution on near-future quantum hardware.

\begin{table}[!h]
\centering
\caption{Logical gate count and depth analysis of the lackadaisical quantum walk search circuit for different grid sizes.}
\renewcommand{\arraystretch}{1.2}
\begin{tabular}{lrrrr}
\toprule
\textbf{Gate} & \textbf{8$\times$8} & \textbf{16$\times$16} & \textbf{32$\times$32} & \textbf{64$\times$64} \\
\midrule
$H$              & 140,250 &   294,709 &   704,663 & 1,607,635 \\
$T$              & 139,531 &   293,452 &   702,746 & 1,607,614 \\
$T^\dagger$      &   3,709 &    12,381 &    42,605 &   130,365 \\
$CX$             &   6,314 &    20,494 &    69,310 &   209,566 \\
$X$              &   1,676 &     4,614 &    15,393 &    47,094 \\
$S$              &  68,793 &   144,311 &    337,923 &   751,017 \\
$Z$              &     291 &       843 &     1,612 &     3,918 \\
$S^\dagger$      &   2,156 &     4,211 &    10,027 &    23,009 \\
Measure          &       6 &         8 &        10 &        12 \\
\midrule
\textbf{Total Gates}        & \textbf{362,726} & \textbf{775,023} & \textbf{1,884,289} & \textbf{4,380,230} \\
\midrule
\textbf{Circuit Depth}      & 213,156 & 457,511 & 1,106,607 & 2,555,755 \\  
\textbf{Non-Clifford Depth} &  73,352 & 179,664 &   357,839 & 1,007,102 \\  
\bottomrule
\end{tabular}
\label{tab:logical}
\end{table}

The component-wise gate counts provide a useful indication of the relative resource requirements of the initialization, oracle, coin, and shift modules; however, their individual costs should not be directly summed to obtain the gate count of the complete LQW circuit. Each component is transpiled independently when its resource cost is evaluated, whereas the complete walk circuit is transpiled as a single composite circuit, allowing the transpiler to optimize across the boundaries between consecutive modules. During this process, adjacent gates may be canceled, combined, or re-synthesized into a more compact sequence, and redundant basis transformations introduced by one component may be removed when followed by operations from the next component. Consequently, the gate count of one complete walk step can differ substantially from the arithmetic sum of the independently evaluated oracle, coin, and shift costs. The same effect becomes more significant for the complete search circuit, where the walk operator is repeated over multiple iterations and additional simplifications can occur between successive steps. Therefore, the component-wise results should be interpreted primarily as a means of identifying which circuit modules contribute most strongly to the resource overhead, while the values reported for the complete circuit represent the actual gate cost after the entire LQW implementation has been jointly transpiled and optimized.

The component-wise resource analysis further shows that different modules of the LQW circuit contribute markedly different amounts to the overall computational cost. For the 8×8 grid, initialization is the least expensive component, with only 414 gates, whereas the standalone oracle and coin require 14,668 and 22,950 gates, respectively, and a single flip-flop shift requires 857 gates. For the 16×16 grid, the initialization cost remains approximately unchanged at 410 gates, and the coin cost also remains approximately constant at 23,04750 gates, while the oracle increases substantially to 45,190 gates and the shift increases to 1,367 gates. This behavior reflects the structure of the implementation: initialization and the lackadaisical coin operate largely on a fixed register structure, particularly the three-qubit coin register, whereas the oracle and shift directly depend on the size of the position registers. As the grid dimension increases, the oracle requires higher-order controlled phase operations and the shift requires more complex controlled increment and decrement circuits, resulting in increased $CX$, $T$, and $T^\dagger$ usage.  The increase in $CX$, $T$, and $T^\dagger$ resources is particularly important because these gates determine much of the entangling and fault-tolerant overhead. Therefore, although the number of algorithmic qubits grows only logarithmically with the number of vertices, the practical resource requirement is governed primarily by the decomposition of multi-controlled oracle operations, controlled position arithmetic, and the repeated non-Clifford operations appearing throughout the search evolution.

\subsection{Fault-tolerant physical resource analysis}
\label{sec:physical-resource-analysis}

The logical resource analysis quantifies the size and depth of the Clifford+$T$ representation of the lackadaisical quantum walk search circuit. These logical quantities, however, cannot be directly interpreted as
fault-tolerant execution resources. In particular, a logical circuit layer does
not in general correspond to a single surface-code cycle \cite{Fowler_2012}. The physical execution
cost depends on the quantum error-correction code distance, the logical
operation implementation, the target failure probability of the full
computation, and the production rate of magic states required by the
non-Clifford operations. We therefore perform the physical resource analysis
using a surface code-based fault-tolerant resource estimation model \cite{vanDam2023QuantumResourceEstimator}. 

For each grid size, the precomputed logical resource counts obtained from the
complete circuit are used as input to the resource estimator. The
circuits are represented in the Clifford+$T$ gate set
\(\{H,X,Z,S,S^\dagger,T,T^\dagger,CX\}\), using the same Qiskit transpilation
settings employed in the logical resource analysis. The estimator input consists
of the algorithmic logical qubit count, the combined \(T/T^\dagger\) count, and
the measurement count. This logical count input follows the standard
precomputed estimate workflow of the Microsoft Quantum Resource Estimator \cite{vanDam2023QuantumResourceEstimator}.  The logical counts describe only the
logical registers of the lackadaisical quantum walk algorithm and do not include
additional fault-tolerant workspace or magic-state factories.

The physical estimates are obtained using the Microsoft Quantum Resource
Estimator with a gate-based superconducting qubit architecture, the predefined
gate-based surface code model, and magic-state distillation for non-Clifford
operations. The baseline architecture assumes a physical operation error rate
\(p=10^{-3}\), a physical gate time of \(100\,\mathrm{ns}\), and a measurement
time of \(500\,\mathrm{ns}\). The total computation failure budget is set to
\(\epsilon=10^{-2}\). The surface code distance is not fixed a priori; it is
selected by the estimator so that the complete computation satisfies the
specified error budget. Similarly, the number of magic-state factories is chosen
according to the non-Clifford resource demand and the imposed space-time
trade-off. The baseline assumptions are summarized in Table~\ref{tab:qre-assumptions}.

\begin{table}[!h]
\centering
\caption{Fault-tolerant resource estimation assumptions used for the
surface-code analysis.}
\label{tab:qre-assumptions}
\begin{tabular}{ll}
\hline
\textbf{Parameter} & \textbf{Value} \\
\hline
Physical architecture & Gate-based superconducting model \\
Physical error rate & \(10^{-3}\) \\
Physical gate time & \(100\,\mathrm{ns}\) \\
Measurement time & \(500\,\mathrm{ns}\) \\
Total error budget & \(10^{-2}\) \\
QEC scheme & Microsoft predefined gate-based surface code \\
Non-Clifford resource & Magic-state distillation \\
Estimator input & Precomputed logical counts \\
Logical circuit basis & Clifford+\(T\) \\
\hline
\end{tabular}
\end{table}

The \(T\) and \(T^\dagger\) gates are the dominant source of
fault-tolerant overhead because each such operation requires a high-fidelity
magic state. From the logical resource analysis, the combined non-Clifford count
increases from 143,240 for the \(8\times 8\) circuit to 1,737,979 for the
\(64\times 64\) circuit. Therefore, the physical cost is determined not only by
the number of logical algorithmic qubits, but also by the rate at which magic
states must be supplied. Allocating more factories can reduce runtime at the
expense of additional physical qubits, while using fewer factories reduces the
spatial footprint, but increases execution time. The physical resource estimate
is therefore a space-time trade-off rather than a single architecture independent
number.

For each grid size, Table~\ref{tab:qre-estimates} reports two representative configurations: (i) the minimum-qubit configuration and (ii) the minimum-runtime configuration. These endpoints illustrate the trade-off between physical qubit footprint and execution time under the specified surface code
model.

\begin{table}[!h]
\centering
\caption{Fault-tolerant physical resource estimates obtained using
the Microsoft Quantum Resource Estimator and the surface-code model. For each
grid size, the minimum-qubit and minimum-runtime configurations
are reported.}
\label{tab:qre-estimates}
\begin{tabular}{llrrrr}
\hline
\textbf{Grid size} & \textbf{Configuration} & \textbf{Code distance} &
\textbf{\(T\) factories} & \textbf{Physical qubits} & \textbf{Runtime} \\
\hline
\(8\times 8\)   & Min. qubits  & 17 & 1  & 34,184  & 41.71 s \\
\(8\times 8\)   & Min. runtime & 15 & 10 & 192,600 & 4.48 s \\
\(16\times 16\) & Min. qubits  & 17 & 1  & 37,074  & 98.48 s \\
\(16\times 16\) & Min. runtime & 15 & 10 & 293,850 & 9.61 s \\
\(32\times 32\) & Min. qubits  & 19 & 1  & 45,436  & 240.00 s \\
\(32\times 32\) & Min. runtime & 17 & 10 & 201,964 & 26.34 s \\
\(64\times 64\) & Min. qubits  & 19 & 1  & 48,324  & 559.63 s \\
\(64\times 64\) & Min. runtime & 17 & 10 & 204,276 & 60.83 s \\
\hline
\end{tabular}
\end{table}

These estimates show that the fault-tolerant resource requirement is governed
primarily by non-Clifford resource production and error-correction overhead.
Although the number of algorithmic logical qubits grows only logarithmically
with the number of grid vertices, the number of logical operations, especially
\(T\)-type operations generated by the decomposed oracle and controlled shift
operations, grows rapidly with grid size. Consequently, larger grids require
larger code distances and substantially longer runtimes.

\section{Conclusion}\label{sec:conclusion}

In this work, we presented a gate-level implementation framework for lackadaisical quantum walk search on two-dimensional grids. The proposed construction realizes the essential components of the algorithm, including position register initialization, the embedding of the five-dimensional lackadaisical coin into a three-qubit unitary representation, phase-oracle construction, and controlled flip-flop shift operations. By translating the abstract walk operator into an executable quantum circuit, this work provides a practical bridge between the theoretical formulation of lackadaisical quantum walks and their implementation on quantum computing platforms. We validated the proposed circuit through ideal noiseless simulations for both single and multiple marked vertices and examined the role of the self-loop weight $\ell$ in determining the search success probability. The results show that the lackadaisical parameter plays a crucial role in controlling amplitude localization at the marked vertex. We further evaluated the circuit under noisy quantum circuit simulations using superconducting device-inspired noise models and applied noise mitigation techniques to improve the observed success probability. In addition, we provided logical and physical resource estimates, including surface-code-based considerations, to assess the scalability of the proposed implementation.

Several directions remain open for future work. First, the circuit construction can be optimized further to reduce gate count, circuit depth, and the number of multi-controlled operations, which would improve its suitability for near-term quantum hardware. Second, the proposed framework can be extended to larger grids, higher-dimensional lattices, irregular graphs, and different configurations of multiple marked vertices. Third, more advanced noise mitigation and error-correction strategies can be incorporated to improve robustness under realistic device constraints. Finally, the implementation framework developed here can support future studies of lackadaisical quantum walks in application domains such as image processing, cryptography, and optimization, where a hardware-executable circuit model is necessary for practical performance evaluation.

\begin{comment}

In this article, we have studied the implementation of the lackadaisical quantum walks on $\sqrt{N}\times\sqrt{N}$ grid for finding one or multiple marked vertices. The experiment was extended to the performance of the circuit in noisy environment and even with noise mitigation methods. Additionally, the physical and logical resources are analyzed in this article. Even the value of $\ell$ (stay probability) was varied during the test under the noisy environment. The mechanism of the circuit in a noisy environment and the optimization of the circuit can be considered as further work.

\end{comment}

\begin{acks}
There is no conflict of interest. The first three authors have contributed equally to this manuscript.
\end{acks}

\bibliographystyle{ACM-Reference-Format}
%\bibliography{sample-base}


%\bibliographystyle{ACM-Reference-Format}
\bibliography{references}
%\newpage
\appendix

\section{Primitive gate description}
\label{app:quantum_gates_matrices}

This section summarizes the primitive quantum gates used in the proposed lackadaisical quantum walk circuit implementation. For each operation, the table provides its circuit-level meaning together with the corresponding matrix representation. These definitions establish the gate notation used throughout the circuit decompositions, including the coin, oracle, and shift-operator constructions.

\begin{table}[H]
\centering
\caption{Primitive quantum gates used in the proposed lackadaisical quantum walk circuit and their matrix representations.}
\small % Slightly smaller text size to fit matrices comfortably
\renewcommand{\arraystretch}{1.5} % Extra vertical spacing for matrices
\begin{tabular}{lll}
\toprule
\textbf{Gate} & \textbf{Description} & \textbf{Matrix Representation} \\
\midrule
\textbf{CX} & Controlled-NOT (Controlled-X) & 
$\begin{pmatrix} 1 & 0 & 0 & 0 \\ 0 & 1 & 0 & 0 \\ 0 & 0 & 0 & 1 \\ 0 & 0 & 1 & 0 \end{pmatrix}$ \\

\textbf{T} & $\pi/8$ Phase Gate ($T$-gate) & 
$\begin{pmatrix} 1 & 0 \\ 0 & e^{i\pi/4} \end{pmatrix}$ \\

\textbf{$\text{T}^\dagger$} & Inverse $T$ Gate ($T$-dagger) & 
$\begin{pmatrix} 1 & 0 \\ 0 & e^{-i\pi/4} \end{pmatrix}$ \\

\textbf{H} & Hadamard Gate & 
$\frac{1}{\sqrt{2}}\begin{pmatrix} 1 & 1 \\ 1 & -1 \end{pmatrix}$ \\

\textbf{S} & Phase Gate ($S$-gate) & 
$\begin{pmatrix} 1 & 0 \\ 0 & i \end{pmatrix}$ \\

\textbf{$\text{S}^\dagger$} & Inverse Phase Gate ($S$-dagger) & 
$\begin{pmatrix} 1 & 0 \\ 0 & -i \end{pmatrix}$ \\

\textbf{X} & Pauli-$X$ Gate (NOT gate) & 
$\begin{pmatrix} 0 & 1 \\ 1 & 0 \end{pmatrix}$ \\

\textbf{Z} & Pauli-$Z$ Gate (Phase-flip gate) & 
$\begin{pmatrix} 1 & 0 \\ 0 & -1 \end{pmatrix}$ \\[3ex]
\bottomrule
\end{tabular}

\label{tab:quantum_gates_matrices}
\end{table}

\end{document}